\documentclass[12pt]{article}

  \usepackage{graphicx}  
  \usepackage{epsfig}
  \usepackage{amssymb}
  \usepackage{subfigure}
    \usepackage{feynmf}
\evensidemargin - 1.truecm
\begin{document}

\begin{fmffile}{QCD_anom_figs}

\newcommand{\be}{\begin{equation}}
\newcommand{\ee}{\end{equation}}
\newcommand{\nn}{\nonumber}
\newcommand{\bea}{\begin{eqnarray}}
\newcommand{\eea}{\end{eqnarray}}
\newcommand{\bfig}{\begin{figure}}
\newcommand{\efig}{\end{figure}}
\newcommand{\bc}{\begin{center}}
\newcommand{\ec}{\end{center}}
\newcommand{\nr}{\hspace{-.4cm}}

\def\gm{\gamma_{\mu}}
\def\g5{\gamma_{5}}
\def\gm5{\gamma_{\mu} \gamma_5}
\def\eps{\epsilon}
\newcommand{\tr}{\mbox{Tr}}

\begin{titlepage}
\nopagebreak
{\flushright{
        \begin{minipage}{5cm}
        PITHA 05/04 \\
        Freiburg-THEP 05/03\\
        ZU-TH 05/05\\
        {\tt hep-ph/0504190}\\
        \end{minipage}        }
}
\vspace*{-1.5cm}                        
\vskip 2.0cm
\begin{center}
{\Large \bf Two-Loop QCD Corrections to the Heavy Quark \\[2mm]
Form Factors: Anomaly Contributions}
\vskip 1.cm
{\large  W.~Bernreuther$\rm \, ^{a, \,}$\footnote{Email: 
{\tt breuther@physik.rwth-aachen.de}}},
{\large  R.~Bonciani$\rm \, ^{b, \,}$\footnote{Email: 
{\tt Roberto.Bonciani@physik.uni-freiburg.de}}},
{\large T.~Gehrmann$\rm \, ^{c,  \, }$\footnote{Email: 
{\tt Thomas.Gehrmann@physik.unizh.ch}}}, \\[2mm] 
{\large R.~Heinesch$\rm \, ^{a, \,}$\footnote{Email: 
{\tt heinesch@physik.rwth-aachen.de}}},
{\large T.~Leineweber$\rm \, ^{a, \,}$\footnote{Email: 
{\tt leineweber@physik.rwth-aachen.de}}}
and {\large E.~Remiddi$\rm \, ^{d, \,}$\footnote{Email: 
{\tt Ettore.Remiddi@bo.infn.it}}}
\vskip .6cm
{\it $\rm ^a$ Institut f\"ur Theoretische Physik, RWTH Aachen,
D-52056 Aachen, Germany} 
\vskip .2cm
{\it $\rm ^b$ Fakult\"at f\"ur Mathematik und Physik, Albert-Ludwigs-Universit\"at
Freiburg, \\ D-79104 Freiburg, Germany} 
\vskip .2cm
{\it $\rm ^c$ Institut f\"ur Theoretische Physik, 
Universit\"at Z\"urich, CH-8057 Z\"urich, Switzerland}
\vskip .2cm
{\it $\rm ^d$ Dipartimento di Fisica dell'Universit\`a di Bologna, and
INFN, Sezione di Bologna, I-40126 Bologna, Italy} 
\end{center}
\vskip .4cm

\begin{abstract}
We present closed
analytic expressions for the order $\alpha_s^2$ triangle diagram contributions to 
the matrix elements  of the singlet and non-singlet axial vector
currents between the vacuum and  a quark-antiquark state. We have
calculated these vertex functions for  arbitrary momentum transfer and
for four different sets of internal and external quark masses.
We show  that both the singlet and non-singlet  vertex functions
satisfy the correct chiral Ward identities. Using the exact
expressions for the finite
axial vector form factors, we check the quality and the convergence of
expansions at production threshold and for asymptotic energies.
\vskip .3cm
\flushright{
        \begin{minipage}{12.3cm}
\hspace*{18.5mm} Heavy quarks, Asymptotic expansion.
        \end{minipage}        }
\end{abstract}
\vfill
\end{titlepage}


%
\section{Introduction \label{sec_intro}}
%

This paper is the third of a series that deals with  the computation
of the  electromagnetic and neutral current form factors of
heavy quarks $Q$ at the two-loop level in QCD
\cite{Bernreuther:2004ih,Bernreuther:2004th}. Knowledge
 of these  form factors puts forward the aim of a
completely differential description of the  electroproduction of heavy
quarks, $e^+ e^- \to \gamma^*, Z^* \to Q {\bar Q} X$,  to order
$\alpha_s^2$ in the QCD coupling. This, in turn, will allow precise
calculations or predictions of many 
observables, including 
forward-backward asymmetries, which are relevant for the
(re)assessment of existing data on $b$ quark production at the $Z$
resonance \cite{unknown:2004qh}, or which will become important for
the production of $b$ quarks and $t$ quarks at a future linear
collider \cite{tesla}. A detailed list of references can be found in 
\cite{Bernreuther:2004ih,Bernreuther:2004th}.
\par
In \cite{Bernreuther:2004ih,Bernreuther:2004th} we presented
analytic results to order $\alpha_s^2$ of the heavy quark
vector and axial vector vertex form factors for arbitrary momentum
transfer, up to  anomalous contributions. Here we close this gap and
analyze the triangle diagram contributions  to the axial vector vertex
which  exhibit the Adler-Bell-Jackiw (ABJ) 
anomaly \cite{Adler:1969gk,Bell:1969ts}. We use dimensional
regularization and the prescription of
\cite{Akyeampong:1973xi,Larin:1993} for implementing the Dirac matrix
$\g5$ in $D\neq 4$, which is an adaptation of the method of \cite{HV}.
In Section 2 we set up the notation and discuss  chiral Ward identities which must be
satisfied by the appropriately 
renormalized axial vector current of a heavy quark, and the
flavor-singlet
and non-singlet axial vector currents, respectively.
In Section 3 we give closed analytic expressions to order $\alpha_s^2$
for the anomalous (pure singlet) contributions to
the axial vector form factors for four different sets of internal
and external quark masses. First we present our results for spacelike
squared momentum transfer $s$; then the  form factors are
analytically continued into the timelike region, and we 
determine also their threshold and asymptotic expansions.
We check that the form factors satisfy the chiral Ward identities
discussed in Section 2. In the context  of this check we have
calculated also the triangle diagram  contributions to the pseudoscalar vertex
function and the vertex function of the gluonic operator $G\tilde G$.
In \cite{Kniehl:1989qu} the triangle contributions to the $b \bar b$ 
vertex function of the non-singlet
current ${\bar t}\gm5 t - {\bar b}\gm5 b$ were computed for massless 
$b$ quarks, using the method of dispersion relations. Our
chirality-conserving form factors  for this specific mass combination
confirm that result. 
In Section 4, we expand our
 exact expressions for the finite 
axial vector form factors at the production threshold and at asymptotic
energies. Detailed comparison of this expansions with the full result
allows us to assess the convergence properties of these widely used expansion
techniques~\cite{hs,smir}.  We conclude in Section 5.
 
%
\section{Chiral Ward Identities \label{sec_chiwa}}
%
We consider QCD with $N_f-1$ massless and one massive quark. 
This includes the case of $N_f=6$ with all quarks but 
the top quark taken to be massless.
We analyze the following external currents:\\
$\bullet$ The   currents 
$J_{5\mu}(x)={\bar Q}(x)\gamma_{\mu}\gamma_5 Q(x)$ and
$J_{5}= {\bar Q}(x)\gamma_5 Q(x)$,  where $Q$ denotes
the field associated with the massive quark.
These currents are called axial vector and  pseudoscalar current in
what follows.  \\
$\bullet$ The  flavor-singlet axial vector current 
$J_{5\mu}^S(x)= \sum {\bar q}(x)\gamma_{\mu}\gamma_5 q(x)$,
where the sum extends over all quark flavors including $Q$. 
 \\
$\bullet$ The  flavor non-singlet neutral axial vector
current  $J_{5\mu}^{NS} =  \sum_{i=1}^3
 {\bar \psi_i}\gamma_{\mu}\gamma_5 \tau_3 \psi_i$,
where $\psi_i = (u,d)_i$ is the i-th  generation quark isodoublet and $\tau_3$ is
the 3rd Pauli matrix. This current couples to the
$Z$ boson.
 The corresponding pseudoscalar current is 
denoted by $J_{5}^{NS}$. \\
The current $J_{5\mu}^{NS}$  is  conserved in the limit of vanishing quark masses,
while the currents $J_{5\mu}$ and $J_{5\mu}^S$ are anomalous at the
quantum level \cite{Adler:1969gk,Bell:1969ts}.

\subsection{Axial Vector and Pseudoscalar Vertex Functions}

In the following we study the matrix elements of the axial vector and
pseudoscalar vertex functions between the vacuum and an on-shell
$q \bar q$ pair to order
$\alpha_s^2$ in the QCD coupling,
 where $q$ denotes one of the $N_f$ quark flavors.
The bare 1-particle irreducible (1PI) vertex functions
involving $J_{5\mu}$ and $J_{5}$
are  denoted by $\Gamma_{q,\mu}$  and $\Gamma_{q}$, respectively.
 The kinematics is $J(p) \to q(p_1) +
{\bar q}(p_2)$, with $p=p_1+p_2$:
\bea
\Gamma_{q,\mu}(s) = \gm5 +\Lambda_{q,\mu}(s)  \, , \\
\Gamma_{q}(s) = \g5 +\Lambda_{q}(s) \; .
\label{g5}
\eea
Following  \cite{Adler:1969gk}
it is useful to distinguish between two types of  contributions to
$\Lambda_{q,\mu}$ and $\Lambda_{q}$ and, likewise, for the other
 vertex functions considered below: those where the  axial vector
 (pseudoscalar)
vertex $\gm5$ $(\g5)$ is attached to the external quark line $q$ are called
type A. For the axial vector and pseudoscalar currents
defined above these are non-zero only for $q=Q$.
Type B contributions  are called those where $\gm5$, respectively $\g5$,
is attached to a  quark loop. To order $\alpha_{s}^2$
these are the triangle diagrams Fig. 1
(a) for the heavy quark currents $J_{5,\mu}$ and $J_5$, and  the
diagrams  Figs. 1 (a,b) 
for $J^S_{5,\mu}$, $J^S_5$,  $J^{NS}_{5,\mu}$, and $J^{NS}_5$.
For  massless quarks in the triangle Fig. 1 (b)
 only Fig. 1 (a) contributes to the matrix elements of
$J_5^S$ and $J_5^{NS}$.
%
%
\bfig[ht]
\bc
\subfigure[]{
\begin{fmfgraph*}(40,25)
\fmfleft{i}
\fmfright{o1,o2}
\fmfright{o1,o2}
\fmf{dashes}{vz,i}
\fmf{phantom}{o1,v1,v2,v3,v4,vz,v5,v6,v7,v8,o2}
\fmffreeze
\fmf{dbl_plain}{vz,v3,v6,vz}
\fmf{plain}{o1,v1,v8,o2}
\fmf{gluon}{v1,v3}
\fmf{gluon}{v6,v8}
\fmflabel{$\bar{q}(p_2)$}{o1}
\fmflabel{$q(p_1)$}{o2}
\end{fmfgraph*}
\label{fig_3a}
}
\quad\quad
\subfigure[]{
\begin{fmfgraph*}(40,25)
\fmfleft{i}
\fmfright{o1,o2}
\fmfright{o1,o2}
\fmf{dashes}{vz,i}
\fmf{phantom}{o1,v1,v2,v3,v4,vz,v5,v6,v7,v8,o2}
\fmffreeze
\fmf{plain}{vz,v6,v3,vz}
\fmf{plain}{o1,v1,v8,o2}
\fmf{gluon}{v1,v3}
\fmf{gluon}{v6,v8}
\fmflabel{$\bar{q}(p_2)$}{o1}
\fmflabel{$q(p_1)$}{o2}
\end{fmfgraph*}
\label{fig_3b}
}
\caption{Triangle diagram (type B) contributions to the
various axial vector and pseudoscalar vertex functions considered in
this paper. The double and single straight line triangles in (a) and (b)
 denote   massive and massless quarks, respectively. The external $q, \,
 \bar q$ quarks are  either massive or massless. The dashed line denotes
 either an axial vector or a pseudoscalar vertex. The crossed diagrams
 are not drawn. \label{fig_3}} 
\ec
\efig

%
%
\bfig[ht]
\bc
\begin{fmfgraph*}(40,25)
\fmfleft{i}
\fmfright{o1,o2}
\fmf{dots}{i,v}
\fmf{phantom}{o1,v1,v2,v3,v,v4,v5,v6,o2}
\fmffreeze
\fmf{plain}{o1,v1,v6,o2}
\fmf{gluon}{v1,v}
\fmf{gluon,rubout}{v,v6}
\fmfdot{v}
\fmflabel{$\bar{q}(p_2)$}{o1}
\fmflabel{$q(p_1)$}{o2}
\end{fmfgraph*}
\caption{One-loop anomalous vertex function $F_q$. \label{fig_1l_ano}}
\ec
\efig
The anomalous Ward identities, which will be discussed in Section 2.3,
involve also the truncated matrix element of the gluonic operator $G{\tilde G}$
between the vacuum and an on-shell $q \bar q$ pair. Here
\be
G(x){\tilde G}(x) \equiv \epsilon_{\mu\nu\alpha\beta}
G^{a,\mu\nu}(x)G^{a,\alpha\beta}(x) \; ,
\label{ggtiop}
\ee
with $G^{a,\mu\nu}$ being the gluon field strength tensor. 
The 
operator (\ref{ggtiop}) induces, to zeroth order in the QCD coupling,
 the momentum-space vertex
$8  k_1^\mu k_2^\alpha
\epsilon_{\mu\nu\alpha\beta}$
with the gluon momentum $k_1$  ($k_2$)
flowing into (out of) the vertex. 
The truncated one-loop $q \bar q$ vertex function $F_q$ which will be
needed in Section 2.3  is shown in Fig. 2.

We use dimensional regularisation in $D=4-2\epsilon$ space-time dimensions. 
For the treatment  of $\g5$ we use a pragmatic approach: the calculation
of the type A contributions is made using  a $\g5$ that anticommutes with $\gamma_\mu$
in $D$ dimensions. This approach works for open fermion lines, but is known to
be algebraically inconsistent, in general, for closed fermion loops with an odd
number of $\g5$ matrices. The appropriate implementation of $\g5$ for type B
diagrams is the prescription of 't Hooft and Veltman \cite{HV} or its adaption
by \cite{Akyeampong:1973xi,Larin:1993} where\footnote{We use the  convention
$\epsilon_{0123} =-\epsilon^{0123} =+1$.}
\be
\g5 = \frac{i}{4!}\epsilon_{\mu_1\mu_2\mu_3\mu_4}\gamma^{\mu_1}\gamma^{\mu_2}\gamma^{\mu_3}\gamma^{\mu_4} \, , \\
\label{dg5}
\ee
\be
\gm5 = \frac{i}{3!}\epsilon_{\mu\mu_1\mu_2\mu_3}\gamma^{\mu_1}\gamma^{\mu_2}\gamma^{\mu_3} \, \\
\label{dgm5}
\ee
is used within dimensional regularisation. 
In these relations, the Lorentz indices
are taken to be $D$-dimensional. The product of two $\epsilon$-tensors is
thus expressed as a determinant over metric tensors in $D$ dimensions.
We employ these relations in this paper. 

In the pragmatic approach chiral invariance is
respected, to second order
in the QCD coupling, also by the unrenormalized 
type A contributions. For the above vertex functions the following
well-known chiral Ward identity can be derived in straightforward fashion:
\be
p^\mu \Lambda_{Q,\mu}^{(A)}  =  2 m_{0,Q} \: \Lambda_{Q}^{(A)}
-\Sigma_Q(p_1)\g5 -\g5 \Sigma_Q(-p_2) \; , \\
\label{wiarep}
\ee
where $m_{0,Q}$ and $\Sigma_Q$ denote the bare mass and self-energy of $Q$.
Because the derivation of ({\ref{wiarep}) was based on an anticommuting 
$\gamma_5$ we have
\be
-\g5 \Sigma_Q(-p_2) =\Sigma_Q(p_2) \g5 \; .
\label{g5s5}
\ee 

\subsection{Renormalization of the Currents}

We renormalize the ultraviolet (UV) divergences  by defining $\alpha_s$ 
in the ${\overline {\rm MS}}$ scheme, while the renormalized quark mass and
the  quark wavefunction renormalization  constant $Z_2$
are defined in the on-shell scheme. That is, the renormalized quark
self-energy is zero on-shell. 
For the on-shell mass  $m_Q$ of the quark  $Q$ we have
 $m_0 = Z_m^{OS} m_Q$, where $Z_m^{OS}$ denotes the mass 
renormalization constant in the on-shell scheme. We described this
hybrid renormalization procedure in detail in 
\cite{Bernreuther:2004ih,Bernreuther:2004th}.

Eq. (\ref{wiarep}) implies that, within the pragmatic approach and
to second order
in the QCD  coupling, neither
the axial vector current $J_{5\mu}$ nor the operator
$m_{0,Q}  {\bar Q}\g5 Q$ 
do get renormalized with respect to type A contributions. 
Thus the renormalized second-order axial vector and pseudoscalar 
vertex functions of type A are given by 
\bea
\Gamma_{Q,\mu}^{(A),R} & = & Z_2^{OS} \, \Gamma_{Q,\mu}^{(A)} \; , \label{gaA}\\
m_Q \, \Gamma_{Q}^{(A),R}&  =  & m_{0,Q} \, Z_2^{OS} \, \Gamma_{Q}^{(A)}=  m_Q
\, Z_m^{OS}  Z_2^{OS} \, \Gamma_{Q}^{(A)} \, \label{gaA5}.
\eea
Nevertheless, the axial vector current must be  renormalized because
of the ABJ anomaly which is present in the triangle diagrams. 
When using an anticommuting $\gamma_5$ for computing
the type A and  (\ref{dg5}), (\ref{dgm5}) for the type B
contributions the correctly
renormalized current to order $\alpha_s^2$ turns out to be
\be 
J_{5\mu}^R = Z_J Z_J^{finite}J_{5\mu} \, ,
\label{pragm1}
\ee
i.e.,  the second order  triangle diagram contributions to
the axial vector vertex function are renormalized according to
\be
\Lambda_{Q,\mu}^{(B),R}  =  \Lambda_{Q,\mu}^{(B)} + (\delta_J +
\delta_5) \gm5  \, , \label{la5mu} 
\ee
where 
\bea 
\delta_J & \equiv & Z_J -1 \, , \nonumber \\
\delta_5 & \equiv & Z_J^{finite} - 1\, .
\eea
The renormalization constant $Z_J$ removes the UV divergences present
in the diagrams Figs. 1 (a,b).
From their computation  -- see the next section --
we find that this 
 constant, which we define in the   $\overline{\rm MS}$ scheme, 
is given by 
\be
Z_J = 1 + \frac{3 C_F T_R}{2 \eps} 
\left(\frac{\alpha_{s}}{2\pi}\right)^2 C^2(\epsilon)+
{\cal O}(\alpha_{s}^3) \, ,
\label{pragm2}
\ee
where $C_F = (N_c^2-1)/2N_c$, $T_R =1/2$, 
$C(\eps)=(4\pi)^{\eps}\Gamma(1+\eps)$, and $N_c$ denotes the number of
colors.
This result is in agreement with  \cite{Larin:1993}. The use of
(\ref{dg5}) and (\ref{dgm5}) 
requires, in addition, a finite renormalization  $Z_J^{finite}$
in order to obtain  the
correct  anomalous Ward  identity  for the axial vector current
\cite{Adler:1969gk}. That is to say, for  the renormalized (singlet)
axial vector currents the non-renormalization \cite{AdlerBa} of the
ABJ anomaly is to be maintained. For the axial vector current this reads
\be
(\partial^\mu J_{5\mu})_R = 2i m_Q ({\bar Q}\g5 Q)_R + \frac{\alpha_{s}}{4\pi}
T_R  (G{\tilde G})_R \; .
\label{abjan}
\ee
From our calculation of the respective vertex functions
involving massive and/or massless quarks,  presented in Section 3, 
we find that   $Z_J^{finite}$ is given by
\be
Z_J^{finite} = 1+ \frac{3 C_F T_R}{4} 
\left(\frac{\alpha_{s}}{2\pi}\right)^2 + {\cal O}(\alpha_{s}^3) \,  . 
\label{pragm3}
\ee
Again, we can compare with the analogous result obtained in
\cite{Larin:1993} for massless QCD, and we find agreement.
 
For a  pseudoscalar vertex the triangle diagrams Fig.~1~(a) are UV and
infrared  finite. This implies that, to order $\alpha_s^2$,
\be
\Lambda_{Q}^{(B),R}  =  \Lambda_{Q}^{(B)} \, ,  
\label{la5m5}
\ee 
and  $m_{0,Q} {\bar Q} \g5 {Q}= 
m_{Q} ({\bar Q} \g5 Q)_R$ within the pragmatic approach.
 We emphasize that no finite renormalization of
the pseudoscalar operator ${\bar Q} \g5 {Q}$ is necessary in order
that the Ward identities 
Eqs. (\ref{wardrama}) and  (\ref{wardsama}) below are satisified. 
In the existing literature,
the issue of a potential finite renormalization of the pseudoscalar singlet
operator was not resolved up to now \cite{Larin:1993}. 
Previous calculations \cite{Chet}
circumvented this issue by forming appropriate non-singlet
combinations  insensitive to this renormalization.

Next we consider the renormalization of the truncated 
$Q \bar Q$ vertex function  $\alpha_s F_Q(s)$ of
the operator $\alpha_s G{\tilde G}$. 
This operator mixes  with $\partial^\mu
J_{5\mu}^S$  under 
renormalization \cite{Espriu:1982bw}:
\be
(G{\tilde G})_R = Z_{G{\tilde G}} \, G{\tilde G} + Z_{GJ} \, \partial^\mu J_{5\mu}^S  \, , 
\label{opmix}
\ee
where $J_{5\mu}^S$ is the flavor-singlet current. The renormalization
constants  which we need here
were given, e.g.,  in \cite{Larin:1993}. In the
  $\overline{\rm MS}$ prescription we have
\bea
Z_{G{\tilde G}} = 1  + {\cal O}(\alpha_{s})  \, , \\
 Z_{GJ} =  \frac{6 C_F}{ \eps} 
\frac{\alpha_{s}}{2\pi} C(\epsilon) +  {\cal O}(\alpha_{s}^2)  \, .
\label{ZGtilde}
\eea
Eqs. (\ref{opmix}) and (\ref{ZGtilde}) imply that the renormalized 
one-loop vertex function $F_Q^R$ is given by
\be
  F_{Q}^R  =   F_Q + iZ_{GJ} \, p^\mu \gm5  \; .
\label{wardradet}
\ee
We recall that  Eq. (\ref{dgm5}) is to be used for  $\gm5$.

Finally we consider the non-singlet currents  $J_{5\mu}^{NS}$ and
$J_{5}^{NS}$, for definiteness
in $N_f=6$ flavor QCD with five massless quarks and the massive quark
$Q$ being the top quark. The divergence $\partial^{\mu}J_{5\mu}^{NS}
= 2m_{0,t} {\bar t}\g5 t$ is non-anomalous, and within our prescription for
implementing $\g5$ the operators $J_{5\mu}^{NS}$ 
and $m_{0,t} {\bar t}\g5 t$  do not get renormalized. 
Let us consider the 1PI vertex functions  involving $J_{5\mu}^{NS}$ and
$J_{5}^{NS}$ and an external on-shell $t \bar t$ pair. Using again the
on-shell scheme for defining the wavefunction and mass of the top
quark, the respective renormalized vertex functions, which contain both type A
and type B contributions, are 
\bea
\Gamma_{t,\mu}^{NS,R} &= &Z_2^{OS} \, \Gamma_{t,\mu}^{NS} \; , \label{tmunons}\\
\Gamma_{t}^{NS,R} & = &  Z_m^{OS} Z_2^{OS} \, \Gamma_{t}^{NS} \,. 
\label{t5nons}
\eea

\subsection{Renormalized Ward Identities}

Let us now discuss the chiral Ward identities for  the various renormalized
vertex functions introduced  in Section 2.2. First to those  of the axial vector and pseudoscalar
currents. For the renormalization scheme specified above 
one derives from Eq. (\ref{wiarep})  that
the type A contributions to the on-shell second order  renormalized vertex functions
(\ref{gaA}), (\ref{gaA5}) must satisfy the Ward identity
 \be
p^\mu \Gamma_{Q,\mu}^{(A),R}  =  2 m_Q \: \Gamma_{Q}^{(A),R} \;  ,
\label{typaR}
\ee
In \cite{Bernreuther:2004th} we computed the 
 vertex function $\Gamma_{Q,\mu}^{(A),R}$ to order $\alpha_s^2$ using
 an anticommuting $\g5$.  
We have checked by explicit computation of  the pseudoscalar
vertex function $\Gamma_{Q}^{(A),R}$, also
to second order in the QCD  coupling, that our results
given in \cite{Bernreuther:2004th} satisfy Eq. (\ref{typaR}).

For the 
second order renormalized type B contributions the anomalous Ward identity
\be 
p^\mu  \Lambda_{Q,\mu}^{(B),R} =
 2 m_Q \Lambda_{Q}^{(B),R}
 -i \frac{\alpha_{s}}{4\pi}  T_R \, F_{Q}^R 
\label{wardrama}
\ee
must hold. This equation follows from Eqs. (\ref{abjan}) 
and (\ref{opmix}). 

With Eq. (\ref{wardrama}) we can immediately write down the Ward
identity for the triangle diagram contributions to the $Q \bar Q$
vertex function of the flavor singlet current $J^S_{5,\mu}$. It reads 
\be 
p^\mu  \sum_{j=1}^{N_f}\Lambda_{Q,\mu}^{(B),R}(s,m_j,m_Q) =
 2 m_Q \Lambda_{Q}^{(B),R}(s,m_Q,m_Q)
 -i \frac{\alpha_{s}}{4\pi} N_f T_R \, F_{Q}^R(s,m_Q) \, ,
\label{wardsama}
\ee
where $m_j$ denotes the mass of the quark in the triangle loop.
All triangle diagrams Figs. 1 (a,b)
contribute to  the left side of (\ref{wardsama}), while for the 
pseudoscalar vertex the Dirac trace over a
triangle is non-zero only for the massive quark. 

Next we consider the vertex functions $\Lambda_{q,\mu}^R$ and  $\Lambda_{q}^R$
which correspond to the matrix elements of $J_{5\mu,R}$ and $J_{5,R}$
between the
vacuum and an on-shell $ q \bar q$ pair with  $q\neq Q$. To second order in the QCD coupling the only
contributions that are non-zero are those of type B. For a massless
on-shell  $ q \bar q$ pair we have 
$\Lambda_{q}^R=0$ to order $\alpha_s^2$ 
due to the chiral invariance of the quark-gluon vertices. The same statement
holds for the vertex function
$F_q^R$ of $(G\tilde G)_R$. Thus for external massless on-shell $q \bar
q$ quarks the corresponding second order renormalized axial vector
vertex function must fulfill
\be
p^\mu  \Lambda_{q,\mu}^{R} = 0 \, .
\label{wardmo}
\ee
The renormalization prescription is that of 
Eqs. (\ref{gaA}) and (\ref{la5mu}). 

Finally  to the non-singlet
vertex functions. Because the operator equation $\partial^{\mu}J_{5\mu}^{NS}
= 2m_{0,t} {\bar t}\g5 t$ does not get renormalized within the
pragmatic approach, the renormalized on-shell vertex functions (\ref{tmunons})
and (\ref{t5nons})  must satisfy the canonical chiral Ward
identity
\be
p^\mu \Gamma_{t,\mu}^{NS,R}  =  2 m_{t} \Gamma_{t}^{NS,R} \;  .
\label{nonsingR}
\ee
The type A and type B contributions to the vertex functions
fulfill Eq. (\ref{nonsingR}) separately. The type B contributions to
$\Gamma_{t,\mu}^{NS}$ and $\Gamma_{t}^{NS}$ are shown in
Figs. 1 (a,b).  Because 
we take massless $u,d,c,s,$ and $b$ quarks, only
the $t$, $b$ quark pair survives the cancellation of
the weak isodoublets in the triangle loops. In fact  $\Gamma_{t}^{NS}
= \Gamma_{t}$ because we have put $m_b=0.$\
If one takes a massless on-shell $q \bar q$ pair instead of $t \bar t$, the
corresponding non-singlet axial vector vertex function must vanish
upon contraction with $p^{\mu}$:
\be
p^\mu \Gamma_{q,\mu}^{NS,R}  = 0 \;  .
\label{qnonsingR}
\ee
In the next section we show by explicit computation  that Eqs.
(\ref{wardrama}), (\ref{wardsama}), (\ref{wardmo}),
 (\ref{nonsingR}),  and  (\ref{qnonsingR}) are satisified by the
 respective type B contributions.

\section{Results for the Triangle Diagrams}

We compute now the contributions Figs. 1 (a,b) to the axial vector and
pseudoscalar vertex functions to order $\alpha_{s}^2$ and the one-loop
anomalous vertex function $F_q$. The on-shell vertex functions are
sandwiched between Dirac spinors 
\bea
\bar{u}(p_1)\Lambda^{(B)}_{q,\mu} \, v(p_2) \, , \\
\bar{u}(p_1)\Lambda^{(B)}_{q} \, v(p_2) \, , \\
\bar{u}(p_1) F_{q} \, v(p_2) \,  ,
\eea
where color indices are suppressed.

\subsection{Calculation}

 Due to parity and CP invariance of QCD
the following decompositions hold in four dimensions:
\bea
\Lambda^{(B)}_{q,\mu}& = & \gm5 \, G_1(s, m_i, m_q) + \frac{1}{2m_q}p^\mu
\g5 \, G_2(s, m_i,m_q) \, , \label{gbmu} \\
\Lambda^{(B)}_{q} & = & A(s, m_i, m_q) \, \g5 \, ,\label{glab5} \\
F_q & = & 2 {i} \, m_q f (s, m_q) \, \g5 \, . \label{gb5}
\eea
The form factors $G_1,$ $G_2,$ $A,$ and  $f$ are dimensionless. 
 As before the mass of the external
quark is denoted by  $m_q$, while $m_i$ is the mass of the quark circulating in the
triangle loop. \\
Because of helicity conservation the form factors $G_2 = A  = F_q =0$
for massless external quarks. Moreover, $A=0$ when the quarks in the
triangle of Fig.~1~(b) are massless. Thus $A\neq 0$ only if both the
internal and external quarks are massive. \\
We compute the renormalized type B two-loop
vertex functions Figs. 1 (a,b) for  the mass combinations $(m_i, m_q)=$  $(m,
m)$, $(0,m)$, $(m,0)$, and $(0,0)$. This is adequate
for a number of applications, for instance to $Z^* \to q \bar q$ and
$Z^* \to Q \bar Q$. Furthermore, we determine  $f(s, m)$.
Let us first consider the axial vector vertex. Its renormalization is
specified in Eq.  (\ref{la5mu}). 
The functions
$\Lambda_{q,\mu}^{(B),R}$
are also infrared (IR) finite for $D \to 4$ 
for any of the above mass combinations. Thus
the renormalized form
factors $G_{j,R}$ are obtained  from these
functions by appropriate projections, which could be carried out,
in principle, in four dimensions using  (\ref{gbmu}). We consider
the spinor traces \cite{Bernreuther:2004th}
\bea
&&\mathcal{P}_{1,R}=\mathrm{Tr}\left[ \gamma_\mu \gamma_5 \, \left( \not\!
    p_1+m_q\right) \left(\Lambda_{q,\mu}^{(B)} + (\delta_J +
\delta_5) \gm5  \right) \left( \not\!
    p_2-m_q\right)\right], \label{proj3}\\
&&\mathcal{P}_{2,R}=\mathrm{Tr}\left[ \gamma_5 \, p_\mu \left( \not\!
    p_1+m_q\right) \left(\Lambda_{q,\mu}^{(B)} + (\delta_J +
\delta_5) \gm5    \right)  \left( \not\!
    p_2-m_q\right)\right] .\label{proj4}
\eea
Rather than calulating $\Lambda_{q,\mu}^{(B),R}$ first and then
performing in (\ref{proj3}),
(\ref{proj4}) the index contractions and traces  
in four dimensions, it is of course technically much simpler to
compute first  the individual, UV-divergent pieces
appearing on the right hand sides
of    (\ref{proj3}) and (\ref{proj4})  in $D$ dimensions.  
Therefore $\gm5$ and $\g5$
appearing in the above traces have to be treated according to 
 (\ref{dg5}), (\ref{dgm5}), even if two $\g5$ appear in the same Dirac
 trace. Because $\mathcal{P}_{1,R}$,
 $\mathcal{P}_{2,R}$ are finite quantities, the relations between them
 and the form factors can be evaluated in four dimensions. We obtain
\be
G_{1,R}(s,m_i,m_q)  =  \phantom{-}
\frac{s \mathcal{P}_{1,R}+2 m_q \mathcal{P}_{2,R}} {4
    s (4 m_q^2 - s )} \, ,
\label{formfactorg1}
\ee
and, if $m_q \neq 0$, 
\be
G_{2,R}(s,m_i,m_q)  = 
- \,{m_q} \frac{m_q s\mathcal{P}_{1,R}+(6 m_q^2- s)\mathcal{P}_{2,R}}
{s^2 (4 m_q^2 -s)} \, .
\label{formfactorg2}
\ee
The form factor associated with the triangle diagram contributions to the 
pseudoscalar vertex function
is both UV- and IR-finite and is given by
\be
A_R(s,m,m)  =  -\frac{1}{2s} \, \tr \left[\gamma_5 \,
  (\not\!p_1+m)\Lambda^{(B)}_{Q}(\not\!p_2-m)\right] \, . 
\label{PAR}
\ee
Eq. (\ref{dg5}) is to be used in the computation of  both the triangle
loop Fig.~1~(a) and of the trace (\ref{PAR}). \\
The anomalous vertex function $F_q$ is nonzero only for a massive
quark. The renormalized form factor, which  is IR-finite, is given by
 \be
f_R(s,m) =  \frac{i}{4ms}
\,\tr\left[\gamma_5 \,  (\not\!p_1+m)\left( F_Q +i  Z_{GJ} p^{\mu} \gm5
     \right)(\not\!p_2-m)\right] \, .
\label{PFR} 
\ee
\par
The above form factors are computed with the technique that was
applied in \cite{Bernreuther:2004ih,Bernreuther:2004th} to the
two-loop vector and type A axial vector form factors. (Cf. these
papers
for  detailed list of references.) Performing the $D$-dimensional
$\gamma$ algebra
using \cite{FORM}
we obtain the form factors expressed in terms of a  set of different
scalar integrals.
These integrals are reduced \cite{Laporta}
to a small set of master integrals.  These
integrals were  evaluated with the method of differential
equations \cite{Rem3,Kot,Rem1,Rem2} in
  \cite{RoPieRem1,RoPieRem2,RoPieRem3}. 
The master integrals and thus the above form factors are
expressed in terms of 1-dimensional harmonic polylogarithms $H(\vec{a};x)$ up to
weight four \cite{Polylog,Polylog3}
 which are functions of the dimensionless variable $x$
defined by Eq. (\ref{xvar}) below.

\subsection{Results for Spacelike $\mathbf s<0$}

We give our results first in the kinematical region in which $s$ is
spacelike. Here the form factors are real. The form
factors involving an internal and/or external mass $m$ are also expressed
in terms of the variable
\be
x = \frac{\sqrt{-s+4m^2} - \sqrt{-s} }{\sqrt{-s+4m^2} + \sqrt{-s}}  \, ,
\label{xvar} 
\ee
with $0 < x < 1$. Defining in four dimensions:
\be
G_{j,R}^{(B)}(s,m_i,m_q) = \left(\frac{\alpha_{s}}{2\pi}\right)^2 C_F
T_R\;\mathcal{G}_j(s,m_i,m_q) \label{gG_i_ano_zerl}
\ee
for $j=1,2$ we obtain for the case $(m,m)$:
\bea
\mathcal{G}_{1}(s,m,m) & = & 3\,\ln(m^2/\mu^2) \, + \Big( \, 32\,x^2\,H(0,0,1,0;x)\,
 -32\,x^2\,H(1,0,0,0;x)\,  \nonumber \\
& &-64\,x^2\,H(0,0,-1,0;x)\, -64\,x^2\,H(0,1,0,0;x)\, \nonumber \\
 & &  +4\,(3\,x^3+x^2-4\,\zeta(2)\,x-3\,x-1)\,x\,H(0,0;x)\, \nonumber \\
& &  +4\,(x^4-2\,x^2-8\,\zeta(2)\,x^2+1)\,H(1,0;x)\, \Big) /\,[(x-1)\,(x+1)^3] \nonumber \\ 
 & &  + \Big( \, 8\,(2\,x^3-3\,x^2+4\,x+1)\,x^2\,H(0,0,0;x)\,
 \nonumber \\
& &  +(3\,x^6-2\,x^5+8\,\zeta(2)\,x^5-8\,\zeta(2)\,x^4-3\,x^4+32\,\zeta(3)\,x^4\nonumber \\
 & &  \phantom{ +(} +32\,\zeta(2)\,x^3 -64\,\zeta(3)\,x^3+4\,x^3+32\,\zeta(3)\,x^2-8\,\zeta(2)\,x^2-3\,x^2\nonumber \\
 & &  \phantom{ +(} -2\,x +8\,\zeta(2)\,x+3)\,H(0;x)\, \Big) /\,[(x+1)^3\,(x-1)^3] \nonumber \\ 
 & &  + 16\,x\,\left( \, H(0,1,0;x)\,    -2 \,H(1,0,0;x)\,
   -2\,H(0,-1,0;x)\, \right) /\,(x+1)^2 \,
 \nonumber \\
 & &  -8\,(x-1)\,H(-1,0;x)\,/\,(x+1) \nonumber \\
 & &  +1/5\,(-45\,x^5+10\,\zeta(2)\,x^5-50\,\zeta(2)\,x^4+80\,\zeta(3)\,x^4-45\,x^4 \nonumber \\ 
 & & \phantom{ +1/5\,(} -80\,\zeta(3)\,x^3+90\,x^3-40\,\zeta(2)\,x^3+16\,\zeta(2)^2\,x^3 \nonumber \\ 
 & & \phantom{ +1/5\,(} -16\,\zeta(2)^2\,x^2-80\,\zeta(3)\,x^2+90\,x^2-40\,\zeta(2)\,x^2-50\,\zeta(2)\,x
\nonumber \\ 
  & & \phantom{ +1/5\,(}
 +80\,\zeta(3)\,x-45\,x-45+10\,\zeta(2))\,/\,[(x-1)^2\,(x+1)^3],
 \label{calg_1_mm}
\eea
where $\mu$ is the mass scale associated with the ${\overline {\rm MS}}$
renormalization of $J_{5,\mu}$ and 
$\zeta(n)$ denotes the Riemann zeta function. The chirality-flipping form factors do not get renormalized. 
For the completely massive case we find:
\bea
\mathcal{G}_{2}(s,m,m) & = & 32\,(2\,x^3-x^2+8\,x+3)\,x^3\,H(0,0,0;x)\,/\,[(x+1)^3\,(x-1)^5] \nonumber \\ 
 & &   -256\,x^3\,H(0,0,-1,0;x)\,/\,[(x-1)^3\,(x+1)^3] \nonumber \\ 
 & &   +64\,x^2\,\big(\, H(0,1,0;x)\,  -2 \,H(1,0,0;x)\, \nonumber \\ 
& & -2 \,H(0,-1,0;x)\, \big) /\,[(x-1)^2\,(x+1)^2] \nonumber \\ 
 & &   -8\,x^2\,\big( H(0,0,0,0;x)\,    + 6 \,H(0,0,0,1;x)\, \big) /\,[(x-1)^3\,(x+1)] \nonumber \\ 
 & &  +32,x^2\, \big(\, H(1,0,0,0;x)\,-\,H(0,0,1,0;x)\,\big)/\,[(x-1)\,(x+1)^3] \nonumber \\ 
 & &   +32\,(x^2-6\,x+1)\,x^2\,H(0,1,0,0;x)\,/\,[(x-1)^3\,(x+1)^3] \nonumber \\ 
 & &   +4\,(3\,x^6+8\,\zeta(2)\,x^5-12\,\zeta(3)\,x^5-2\,x^5+8\,\zeta(2)\,x^4-3\,x^4 \nonumber \\ 
 & & \phantom{  +4\,(} +32\,\zeta(3)\,x^4-40\,\zeta(3)\,x^3+4\,x^3+64\,\zeta(2)\,x^3-3\,x^2 \nonumber \\ 
 & & \phantom{  +4\,(} +32\,\zeta(3)\,x^2+8\,\zeta(2)\,x^2-12\,\zeta(3)\,x+8\,\zeta(2)\,x-2\,x \nonumber \\ 
 & & \phantom{  +4\,(} +3)\,x\,H(0;x)\,/\,[(x+1)^3\,(x-1)^5] \nonumber \\ 
 & &   +4\,(13\,x^5-13\,x^4-4\,\zeta(2)\,x^4-30\,x^3-20\,\zeta(2)\,x^3+20\,\zeta(2)\,x^2 \nonumber \\ 
 & & \phantom{  +4\,(} -2\,x^2+4\,\zeta(2)\,x+x-1)\,x\,H(0,0;x)\,/\,[(x-1)^4\,(x+1)^3] \nonumber \\ 
 & &   +16\,(x^2+2\,\zeta(2)\,x+2\,x+1)\,x\,H(1,0;x)\,/\,[(x-1)\,(x+1)^3] \nonumber \\ 
 & &   +8\,x\,\big( H(0,1;x)\, -4 \, H(-1,0;x)\, \big) /\,[(x-1)\,(x+1)] \nonumber \\ 
 & &   +8/5\,\Big(15\,\zeta(2)\,x^5-5\,x^5-8\,\zeta(2)^2\,x^4-55\,\zeta(2)\,x^4+40\,\zeta(3)\,x^4 \nonumber \\ 
 & & \phantom{  +8/5\,(} -5\,x^4+10\,x^3-100\,\zeta(2)\,x^3-40\,\zeta(3)\,x^3-60\,\zeta(2)\,x^2 \nonumber \\ 
 & & \phantom{  +8/5\,(} +10\,x^2-40\,\zeta(3)\,x^2+8\,\zeta(2)^2\,x-5\,x-35\,\zeta(2)\,x \nonumber \\ 
 & & \phantom{  +8/5\,(}
 +40\,\zeta(3)\,x-5\,\zeta(2)-5 \Big)\,x\,/\,[(x-1)^4\,(x+1)^3].\label{calg_2_mm}
\eea
For the case of massless internal quarks we obtain:
\bea
\mathcal{G}_{1}(s,0,m) & = & 3\,\ln(m^2/\mu^2) \,+ 8\,x^2\,H(0,1,0;x)\,/\,[(x+1)^3\,(x-1)] \nonumber \\ 
 & &  +4\,x^2\,H(0,0,0;x)\,/\,[(x+1)^3\,(x-1)] \nonumber \\ 
 & &  +8\,x^2\,H(0,0,1;x)\,/\,[(x+1)^3\,(x-1)] \nonumber \\ 
 & &  +16\,x^2\,H(0,1,1;x)\,/\,[(x+1)^3\,(x-1)] \nonumber \\ 
 & &  -2\,(x^2+3\,x-2)\,x\,H(0,0;x)\,/\,(x+1)^3 \nonumber \\
 & &  -4\,(x^2+3\,x-2)\,x\,H(0,1;x)\,/\,(x+1)^3 \nonumber \\
 & &  -2\,(3\,x^2-2\,x+3)\,H(1;x)\,/\,(x+1)^2 \nonumber \\
 & &  -2\,(x^2+1)\,H(1,0;x)\,/\,(x+1)^2 \nonumber \\
 & &  +(-3\,x^4+2\,x^3+8\,\zeta(2)\,x^2-2\,x+3)\,H(0;x)\,/\,[(x+1)^3\,(x-1)] \nonumber \\ 
 & &  -4\,(x^2+1)\,H(1,1;x)\,/\,(x+1)^2 \nonumber \\
 & &  -(9\,x^4+6\,\zeta(2)\,x^4+18\,x^3+16\,\zeta(2)\,x^3+8\,\zeta(3)\,x^2-40\,\zeta(2)\,x^2 \nonumber \\ 
 & & \phantom{ -(}
 -18\,x+16\,\zeta(2)\,x-9+2\,\zeta(2))\,/\,[(x+1)^3\,(x-1)] \, , 
\label{calg_1_0m}
\eea
and
\bea
\mathcal{G}_{2}(s,0,m) & = & 32\,x^3\,H(0,1,0;x)\,/\,[(x-1)^3\,(x+1)^3] \nonumber \\ 
 & &  +16\,x^3\,H(0,0,0;x)\,/\,[(x-1)^3\,(x+1)^3] \nonumber \\ 
 & &  +32\,x^3\,H(0,0,1;x)\,/\,[(x-1)^3\,(x+1)^3] \nonumber \\ 
 & &  +64\,x^3\,H(0,1,1;x)\,/\,[(x-1)^3\,(x+1)^3] \nonumber \\ 
 & &  -4\,(x^3+5\,x^2-3\,x+1)\,x\,H(0,0;x)\,/\,[(x+1)^3\,(x-1)^2] \nonumber \\ 
 & &  -8\,(3\,x^2-2\,x+3)\,x\,H(1;x)\,/\,[(x+1)^2\,(x-1)^2] \nonumber \\ 
 & &  -8\,(x^3+5\,x^2-3\,x+1)\,x\,H(0,1;x)\,/\,[(x+1)^3\,(x-1)^2] \nonumber \\ 
 & &  +4\,(-3\,x^4+2\,x^3+8\,\zeta(2)\,x^2-2\,x+3)\,x\,H(0;x) \nonumber \\ 
 & & \phantom{ +4\,(} /\,[(x-1)^3\,(x+1)^3] \nonumber \\ 
 & &  -8\,(x^2+1)\,x\,H(1,0;x)\,/\,[(x+1)^2\,(x-1)^2] \nonumber \\ 
 & &  -16\,(x^2+1)\,x\,H(1,1;x)\,/\,[(x+1)^2\,(x-1)^2] \nonumber \\ 
 & &  -8\,(x^4+\zeta(2)\,x^4+2\,x^3+8\,\zeta(2)\,x^3-16\,\zeta(2)\,x^2+4\,\zeta(3)\,x^2 \nonumber \\ 
 & & \phantom{ -8\,(}
 +8\,\zeta(2)\,x-2\,x-\zeta(2)-1)\,x\,/\,[(x-1)^3\,(x+1)^3].
 \label{calg_2_0m}
\eea
For the case of massless external quarks we get:
\bea
\mathcal{G}_{1}(s,m,0) & = &   3\,\ln(-s/\mu^2) \, + 2\,(5\,x^2-8\,x+7)\,x^2\,H(0,0;x)\,/\,(x-1)^4 \nonumber \\
 & &  -8\,x^2\,H(0,1,0;x)\,/\,(x-1)^4 \nonumber \\
 & &  -4\,x^2\,H(0,0,0;x)\,/\,(x-1)^4 \nonumber \\
 & &  -2\,(-3\,x^3+4\,\zeta(2)\,x^2+9\,x^2-9\,x-4\,\zeta(2)\,x \nonumber \\ 
 & & \phantom{ -2\,(} +3+4\,\zeta(2))\,x\,H(0;x)\,/\,(x-1)^4 \nonumber \\
 & &  -16\,x\,H(1,0,0;x)\,/\,(x-1)^2 \nonumber \\
 & &  -16\,x\,H(0,-1,0;x)\,/\,(x-1)^2 \nonumber \\
 & &  +2\,(x+1)\,(x^2+1)\,H(1,0;x)\,/\,(x-1)^3 \nonumber \\
 & &  +6\,H(1;x)-8\,(x+1)\,H(-1,0;x)\,/\,(x-1) \nonumber \\
 & &  -(2\,\zeta(2)\,x^4+9\,x^4+8\,\zeta(3)\,x^3-8\,\zeta(2)\,x^3-36\,x^3\nonumber \\ 
 & & \phantom{ -(} +54\,x^2-36\,x +8\,\zeta(2)\,x+8\,\zeta(3)\,x+9 \nonumber \\ 
 & & \phantom{ -(} -2\,\zeta(2))\,/\,(x-1)^4 \, , 
\label{g1m0}
\eea
and
\bea
\mathcal{G}_{1}(s,0,0) & = &  3\,\ln(-s/\mu^2) \, -9+2\,\zeta(2)\, .
\label{g100}
\eea
Moreover, we have 
\be 
\mathcal{G}_{2}(s,m,0) = \mathcal{G}_{2}(s,0,0) = 0 \, .
\ee
As already mentioned above the form factor associated with the pseudoscalar
vertex function is non-zero only for the completely massive case. It is given by
\bea
 & & A_R(s,m,m) =  \left(\frac{\alpha_{s}}{2\pi}\right)^2 C_F T_R \,
 \Big(\, -16\,x^2\,H(0,0,0;x)\,/\,[(x-1)^3\,(x+1)] \nonumber \\
 & & \hspace{1cm}  +4\,(x+\zeta(2)\,x-\zeta(2)+1)\,x\,H(0,0;x)\,/\,[(x+1)\,(x-1)^2] \nonumber \\ 
 & & \hspace{1cm}  +4\,(3\,\zeta(3)\,x^2-4\,\zeta(2)\,x-6\,\zeta(3)\,x+3\,\zeta(3))\,x\,H(0;x)\, \nonumber \\ 
 & & \hspace{1cm} \phantom{ +4\,(} /\,[(x-1)^3\,(x+1)] \nonumber \\ 
 & & \hspace{1cm}  +12\,x\,H(0,0,0,1;x)\,/\,[(x+1)\,(x-1)] \nonumber \\ 
 & & \hspace{1cm}  -8\,\zeta(2)\,x\,H(1,0;x)\,/\,[(x+1)\,(x-1)] \nonumber \\ 
 & & \hspace{1cm}  +2\,x\,H(0,0,0,0;x)\,/\,[(x+1)\,(x-1)] \nonumber \\ 
 & & \hspace{1cm}  +8\,x\,H(0,0,1,0;x)\,/\,[(x+1)\,(x-1)] \nonumber \\ 
 & & \hspace{1cm}  -8\,x\,H(0,1,0,0;x)\,/\,[(x+1)\,(x-1)] \nonumber \\ 
 & & \hspace{1cm}  -8\,x\,H(1,0,0,0;x)\,/\,[(x+1)\,(x-1)] \nonumber \\ 
 & & \hspace{1cm} 
 +8/5\,(2\,\zeta(2)\,x+5\,x-2\,\zeta(2)+5)\,\zeta(2)\,x\,/\,[(x+1)\,(x-1)^2]\Big) \,. 
\label{Afofa}
\eea
Finally the form factor associated with the anomalous vertex function Fig. 2
is:
\bea
  f_{R}(s,m) = 
 \left(\frac{\alpha_{s}}{2\pi}\right)  \, 2\, C_F  \,  
 \Big(\, 3  \,\ln(m^2/\mu^2) \,  -(x-1)\,H(0,0;x)\,/\,(x+1) \nonumber \\ 
  -2\,(x-1)\,H(0,1;x)\,/\,(x+1) \,
 -(7\,x+4\,\zeta(2)\,x+7-4\,\zeta(2))\,/\,(x+1)\Big). 
\label{anofr}
\eea

\subsection{Analytical Continuation above Threshold\label{subsec_analy}}

We
perform the analytical continuation of the form factors
to the physical region above
threshold, $s > 4m^2$, $-1 < x < 0$, respectively
$s > 0$ for massless external quarks, 
with the usual 
prescription  $s \to s +i \delta$. 
For the cases involving masses the variable $x$ becomes a phase
factor if  $0< s <4m^2$ \cite{Bernreuther:2004ih}. For  $s>4m^2$ we define:
\be
y = \frac{\sqrt{s} - \sqrt{s-4m^2} }{\sqrt{s} + \sqrt{s-4m^2} } \, ,
\ee
and the continuation in $x$ is performed by the replacement:
\be
x \rightarrow - y + i \delta \, .
\ee
The real and imaginary parts of the form factors are defined through the
relations:
\be
G_{j,R}(s+i\delta)  =  {\rm Re} \, G_{j,R}(s) 
                         + i \pi \, {\rm Im }\, G_{j,R}(s) \,  , 
\ee
and likewise for $A_R$ and $f_R$,  where also a factor $\pi$ is taken
out of the respective imaginary part. \\
In the completely massive case we get for $s> 4m^2$:
\bea
\mathrm{Re} \, \mathcal{G}_{1}(s,m,m)  & = & -8\,(2\,y^3+3\,y^2+4\,y-1)\,y^2\,H(0,0,0;y)\,/\,[(y-1)^3\,(y+1)^3] \nonumber \\ 
 & &  +32\,y^2\,H(-1,0,0,0;y)\,/\,[(y+1)\,(y-1)^3] \nonumber \\ 
 & &  +64\,y^2\,H(0,-1,0,0;y)\,/\,[(y+1)\,(y-1)^3] \nonumber \\ 
 & &  -192\,\zeta(2)\,y^2\,H(0,-1;y)\,/\,[(y+1)\,(y-1)^3] \nonumber \\ 
 & &  -32\,y^2\,H(0,0,-1,0;y)\,/\,[(y+1)\,(y-1)^3] \nonumber \\ 
 & &  +64\,y^2\,H(0,0,1,0;y)\,/\,[(y+1)\,(y-1)^3] \nonumber \\ 
 & &  -32\,y\,H(0,1,0;y)\,/\,(y-1)^2 \nonumber \\ 
 & &  +16\,y\,H(0,-1,0;y)\,/\,(y-1)^2 \nonumber \\ 
 & &  +4\,(3\,y^3-y^2-3\,y-4\,\zeta(2)\,y+1)\,y\,H(0,0;y)\, \nonumber \\ 
 & & \phantom{ +4\,(} /\,[(y+1)\,(y-1)^3] \nonumber \\ 
 & &  -32\,y\,H(-1,0,0;y)\,/\,(y-1)^2 \nonumber \\ 
 & &  +96\,\zeta(2)\,y\,H(-1;y)\,/\,(y-1)^2 \nonumber \\ 
 & &  +(3\,y^6+2\,y^5+40\,\zeta(2)\,y^5+32\,\zeta(3)\,y^4+64\,\zeta(2)\,y^4-3\,y^4 \nonumber \\ 
 & & \phantom{ +(} +64\,\zeta(2)\,y^3+64\,\zeta(3)\,y^3-4\,y^3+32\,\zeta(3)\,y^2-32\,\zeta(2)\,y^2 \nonumber \\ 
 & & \phantom{ +(} -3\,y^2+2\,y-8\,\zeta(2)\,y+3)\,H(0;y)\,/\,[(y-1)^3\,(y+1)^3] \nonumber \\ 
 & &  +8\,(y+1)\,H(1,0;y)\,/\,(y-1) \nonumber \\ 
 & &  -4\,(y^4-2\,y^2+16\,\zeta(2)\,y^2+1)\,H(-1,0;y)\,/\,[(y+1)\,(y-1)^3] \nonumber \\ 
 & &  -1/5\,(45\,y^5+170\,\zeta(2)\,y^5+80\,\zeta(3)\,y^4+70\,\zeta(2)\,y^4-45\,y^4\nonumber \\ 
 & & \phantom{ -1/5\,(} -90\,y^3-256\,\zeta(2)^2\,y^3-200\,\zeta(2)\,y^3+80\,\zeta(3)\,y^3 \nonumber \\ 
 & & \phantom{ -1/5\,(} -160\,\zeta(2)\,y^2-256\,\zeta(2)^2\,y^2+90\,y^2-80\,\zeta(3)\,y^2 \nonumber \\ 
 & & \phantom{ -1/5\,(} -80\,\zeta(3)\,y+45\,y+110\,\zeta(2)\,y-45+10\,\zeta(2))\, \nonumber \\ 
 & & \phantom{ -1/5\,(} /\,[(y+1)^2\,(y-1)^3]\nonumber \\
 & & +3\,\ln(m^2/\mu^2), 
\eea
and
\bea
\mathrm{Im} \, \mathcal{G}_{1}(s,m,m)  & = & -8\,(2\,y^3+3\,y^2+4\,y-1)\,y^2\,H(0,0;y)\,/\,[(y-1)^3\,(y+1)^3] \nonumber \\ 
 & &  +32\,y^2\,H(-1,0,0;y)\,/\,[(y+1)\,(y-1)^3] \nonumber \\ 
 & &  +64\,y^2\,H(0,-1,0;y)\,/\,[(y+1)\,(y-1)^3] \nonumber \\ 
 & &  -32\,y^2\,H(0,0,-1;y)\,/\,[(y+1)\,(y-1)^3] \nonumber \\ 
 & &  +64\,y^2\,H(0,0,1;y)\,/\,[(y+1)\,(y-1)^3] \nonumber \\ 
 & &  +4\,(3\,y^3-y^2-4\,\zeta(2)\,y-3\,y+1)\,y\,H(0;y)\,/\,[(y+1)\,(y-1)^3] \nonumber \\ 
 & &  -32\,y\,H(-1,0;y)\,/\,(y-1)^2 \nonumber \\ 
 & &  +16\,y\,H(0,-1;y)\,/\,(y-1)^2 \nonumber \\ 
 & &  -32\,y\,H(0,1;y)\,/\,(y-1)^2 \nonumber \\ 
 & &  -4\,(y+1)\,H(-1;y)\,/\,(y-1) \nonumber \\ 
 & &  +8\,(y+1)\,H(1;y)\,/\,(y-1) \nonumber \\ 
 & &  +(3\,y^4-4\,y^3+8\,\zeta(2)\,y^3+2\,y^2+32\,\zeta(3)\,y^2-4\,y-8\,\zeta(2)\,y \nonumber \\ 
 & & \phantom{ +(} +3)\,/\,[(y+1)\,(y-1)^3].
\eea
For the chirality-flipping form factor we have
\bea
\mathrm{Re} \, \mathcal{G}_{2}(s,m,m) & = &  32\,(2\,y^3+y^2+8\,y-3)\,y^3\,H(0,0,0;y)\,/\,[(y-1)^3\,(y+1)^5] \nonumber \\ 
 & &  -256\,y^3\,H(0,0,1,0;y)\,/\,[(y+1)^3\,(y-1)^3] \nonumber \\ 
 & &  -384\,\zeta(2)\,y^2\,H(-1;y)\,/\,[(y+1)^2\,(y-1)^2] \nonumber \\ 
 & &  -8\,y^2\,H(0,0,0,0;y)\,/\,[(y+1)^3\,(y-1)] \nonumber \\ 
 & &  -32\,(y^2+6\,y+1)\,y^2\,H(0,-1,0,0;y)\,/\,[(y+1)^3\,(y-1)^3] \nonumber \\ 
 & &  +128\,y^2\,H(-1,0,0;y)\,/\,[(y+1)^2\,(y-1)^2] \nonumber \\ 
 & &  -32\,y^2\,H(-1,0,0,0;y)\,/\,[(y+1)\,(y-1)^3] \nonumber \\ 
 & &  -64\,y^2\,H(0,-1,0;y)\,/\,[(y+1)^2\,(y-1)^2] \nonumber \\ 
 & &  +128\,y^2\,H(0,1,0;y)\,/\,[(y+1)^2\,(y-1)^2] \nonumber \\ 
 & &  +32\,y^2\,H(0,0,-1,0;y)\,/\,[(y+1)\,(y-1)^3] \nonumber \\ 
 & &  +48\,y^2\,H(0,0,0,-1;y)\,/\,[(y+1)^3\,(y-1)] \nonumber \\ 
 & &  -4\,(13\,y^5-2\,\zeta(2)\,y^4+13\,y^4-30\,y^3-14\,\zeta(2)\,y^3-14\,\zeta(2)\,y^2 \nonumber \\ 
 & & \phantom{ -4\,(} +2\,y^2-2\,\zeta(2)\,y+y+1)\,y\,H(0,0;y)\,/\,[(y+1)^4\,(y-1)^3] \nonumber \\ 
 & &  +8\,(y^4+12\,\zeta(2)\,y^3-2\,y^2+72\,\zeta(2)\,y^2+12\,\zeta(2)\,y \nonumber \\ 
 & & \phantom{ +8\,(} +1)\,y\,H(0,-1;y)\,/\,[(y+1)^3\,(y-1)^3] \nonumber \\ 
 & &  +16\,(y^2+4\,\zeta(2)\,y-2\,y+1)\,y\,H(-1,0;y)\,/\,[(y+1)\,(y-1)^3] \nonumber \\ 
 & &  -4\,(3\,y^6+12\,\zeta(3)\,y^5+40\,\zeta(2)\,y^5+2\,y^5+32\,\zeta(2)\,y^4\nonumber \\ 
 & & \phantom{ -4\,(} +32\,\zeta(3)\,y^4 -3\,y^4+40\,\zeta(3)\,y^3-4\,y^3+128\,\zeta(2)\,y^3 \nonumber \\ 
 & & \phantom{ -4\,(} +32\,\zeta(3)\,y^2-3\,y^2-64\,\zeta(2)\,y^2+2\,y+12\,\zeta(3)\,y \nonumber \\ 
 & & \phantom{ -4\,(} -8\,\zeta(2)\,y+3)\,y\,H(0;y)\,/\,[(y-1)^3\,(y+1)^5] \nonumber \\ 
 & &  -32\,y\,H(1,0;y)\,/\,[(y+1)\,(y-1)] \nonumber \\ 
 & &  +4/5\,(10\,y^4+165\,\zeta(2)\,y^4+80\,\zeta(3)\,y^3+29\,\zeta(2)^2\,y^3\nonumber \\ 
 & & \phantom{ +4/5\,(} -80\,\zeta(2)\,y^3-20\,y^3-314\,\zeta(2)^2\,y^2-170\,\zeta(2)\,y^2 \nonumber \\ 
 & & \phantom{ +4/5\,(} +29\,\zeta(2)^2\,y+80\,\zeta(2)\,y+20\,y-80\,\zeta(3)\,y-10 \nonumber \\ 
 & & \phantom{ +4/5\,(}
 +5\,\zeta(2))\,y\,/\,[(y+1)^3\,(y-1)^3],
\eea
and
\bea
\mathrm{Im}\,\mathcal{G}_{2}(s,m,m) & = & -256\,y^3\,H(0,0,1;y)\,/\,[(y+1)^3\,(y-1)^3] \nonumber \\ 
 & &  +32\,(2\,y^3+y^2+8\,y-3)\,y^3\,H(0,0;y)\,/\,[(y-1)^3\,(y+1)^5] \nonumber \\ 
 & &  +128\,y^2\,H(0,1;y)\,/\,[(y+1)^2\,(y-1)^2] \nonumber \\ 
 & &  +32\,y^2\,H(0,0,-1;y)\,/\,[(y+1)\,(y-1)^3] \nonumber \\ 
 & &  -64\,y^2\,H(0,-1;y)\,/\,[(y+1)^2\,(y-1)^2] \nonumber \\ 
 & &  -32\,(y^2+6\,y+1)\,y^2\,H(0,-1,0;y)\,/\,[(y+1)^3\,(y-1)^3] \nonumber \\ 
 & &  +128\,y^2\,H(-1,0;y)\,/\,[(y+1)^2\,(y-1)^2] \nonumber \\ 
 & &  -8\,y^2\,H(0,0,0;y)\,/\,[(y+1)^3\,(y-1)] \nonumber \\ 
 & &  -32\,y^2\,H(-1,0,0;y)\,/\,[(y+1)\,(y-1)^3] \nonumber \\ 
 & &  -32\,y\,H(1;y)\,/\,[(y+1)\,(y-1)] \nonumber \\ 
 & &  +16\,y\,H(-1;y)\,/\,[(y+1)\,(y-1)] \nonumber \\ 
 & &  -4\,(13\,y^5+13\,y^4+2\,\zeta(2)\,y^4-30\,y^3-18\,\zeta(2)\,y^3+2\,y^2 \nonumber \\ 
 & & \phantom{ -4\,(} -18\,\zeta(2)\,y^2+y+2\,\zeta(2)\,y+1)\,y\,H(0;y)\, \nonumber \\ 
 & & \phantom{ -4\,(} /\,[(y+1)^4\,(y-1)^3] \nonumber \\ 
 & &  -4\,(3\,y^4+12\,\zeta(3)\,y^3-4\,y^3+8\,\zeta(2)\,y^3+2\,y^2+8\,\zeta(3)\,y^2 \nonumber \\ 
 & & \phantom{ -4\,(}
 -4\,y+12\,\zeta(3)\,y-8\,\zeta(2)\,y+3)\,y\,/\,[(y+1)^3\,(y-1)^3].\nonumber \\
\eea 
We obtain for the non-zero pseudoscalar and anomaly form factors:
\bea
\mathrm{Re} \, {A}_{R}(s,m,m) & =  & \left(\frac{\alpha_{s}}{2\pi}\right)^2 \, 
\frac{C_F\,T_R\;y}{(1-y)(1+y)^3}\;\Big[4\, ( 3\,\zeta(3) +3\,\zeta(3) {y}^{2}+6\,y\zeta(3) -8\,y\zeta(2)) H(0;y) \nn\\
& & -2\, (1+y)(2-2\,y+y\zeta(2) +\zeta(2)) H(0,0;y) +16\,yH(0,0,0;y) \nn\\
& &  -12\, (1+y)^{2}H(0,0,0,-1;y) +2\,(1+y)^{2}H(0,0,0,0;y) \nn\\
& & -8\, (1+y)^{2}H(0,0,-1,0;y) -24\,\zeta(2)(1+y)^{2}H(0,-1;y) \nn\\
& & +8\, (1+y)^{2}H(0,-1,0,0;y) -16\,\zeta(2)(1+y)^{2}H(-1,0;y) \nn\\
& & +8\, (1+y)^{2}H(-1,0,0,0;y) -{\frac{58}{5}}\,y
\zeta^2(2)-4\,{y}^{2}\zeta(2) \nn \\
& & -{\frac{29}{5}}\,\zeta^2(2) 
+4\,\zeta(2) -{\frac{29}{5}}\,\zeta^2(2){y}^{2}\Big] \, ,
\eea
\bea
\mathrm{Im}\, {A}_{R}(s,m,m) & = & \left(\frac{\alpha_{s}}{2\pi}\right)^2 \,
\frac{C_F\,T_R\;y}{(1-y)(1+y)^3}\;\Big[2\,(1+y)(
y\zeta(2)+2\,y-2+\zeta(2))H(0;y) \nn \\
& & +16\,yH(0,0;y) \,  +2\,(1+y)^{2}H(0,0,0;y)
-8\,(1+y)^{2}H(0,0,-1;y) \nn \\
& & +8\,(1+y)^{2}H(0,-1,0;y) \nn\\
& & +8\,(1+y)^{2}H(-1,0,0;y) +24\,y\zeta(3) +12\,\zeta(3)
{y}^{2}+12\,\zeta(3)\Big] \, ,
\eea
\bea
\mathrm{Re} \, {f}_{R}(s,m) & = &  \left(\frac{\alpha_{s}}{2\pi}\right)
\frac{2 \,C_F}{1-y}\,\Big[( 1+y ) \,H(0,0;y) -2( 1+y )\,H(0,-1;y)
+\zeta(2)(1+y) \nonumber \\
& & +3\,(1-y) \ln (m^2/\mu^2) -7(1-y)\Big] \, , \\
\mathrm{Im} \,  {f}_{R}(s,m) & = &
\left(\frac{\alpha_{s}}{2\pi}\right) \frac{2
  \,C_F}{1-y}( 1+y )\, H(0;y)   \, .
\eea

In the case of  massless internal quarks we get for $s > 4m^2$:
\bea
\mathrm{Re} \, \mathcal{G}_{1}(s,0,m) & = & -8\,y^2\,H(0,0,-1;y)\,/\,[(y-1)^3\,(y+1)] \nonumber \\ 
 & &  +4\,y^2\,H(0,0,0;y)\,/\,[(y-1)^3\,(y+1)] \nonumber \\ 
 & &  -8\,y^2\,H(0,-1,0;y)\,/\,[(y-1)^3\,(y+1)] \nonumber \\ 
 & &  +16\,y^2\,H(0,-1,-1;y)\,/\,[(y-1)^3\,(y+1)] \nonumber \\ 
 & &  +4\,(y^2-3\,y-2)\,y\,H(0,-1;y)\,/\,(y-1)^3 \nonumber \\ 
 & &  -2\,(y^2-3\,y-2)\,y\,H(0,0;y)\,/\,(y-1)^3 \nonumber \\ 
 & &  -(3\,y^4+2\,y^3+4\,\zeta(2)\,y^2-2\,y-3)\,H(0;y)\, \nonumber \\ 
 & & \phantom{ -(} /\,[(y-1)^3\,(y+1)] \nonumber \\ 
 & &  +2\,(3\,y^2+2\,y+3)\,H(-1;y)\,/\,(y-1)^2 \nonumber \\ 
 & &  -4\,(y^2+1)\,H(-1,-1;y)\,/\,(y-1)^2 \nonumber \\ 
 & &  +2\,(y^2+1)\,H(-1,0;y)\,/\,(y-1)^2 \nonumber \\ 
 & &  -(9\,y^4-18\,y^3-4\,\zeta(2)\,y^3-10\,\zeta(2)\,y^2+8\,\zeta(3)\,y^2+18\,y \nonumber \\ 
 & & \phantom{ -(}
 -4\,\zeta(2)\,y+2\,\zeta(2)-9)\,/\,[(y-1)^3\,(y+1)]\nonumber \\
 & & +3\,\ln(m^2/\mu^2),
\eea
\bea
\mathrm{Im} \, \mathcal{G}_{1}(s,0,m) & = & -8\,y^2\,H(0,-1;y)\,/\,[(y-1)^3\,(y+1)] \nonumber \\ 
 & &  +4\,y^2\,H(0,0;y)\,/\,[(y-1)^3\,(y+1)] \nonumber \\ 
 & &  -2\,(y^2-3\,y-2)\,y\,H(0;y)\,/\,(y-1)^3 \nonumber \\ 
 & &  +2\,(y^2+1)\,H(-1;y)\,/\,(y-1)^2 \nonumber \\ 
 & &
 -(3\,y^4+2\,y^3-4\,\zeta(2)\,y^2-2\,y-3)\,/\,[(y-1)^3\,(y+1)],
\eea
\bea
\mathrm{Re} \, \mathcal{G}_{2}(s,0,m) & = & 32\,y^3\,H(0,-1,0;y)\,/\,[(y+1)^3\,(y-1)^3] \nonumber \\ 
 & &  +32\,y^3\,H(0,0,-1;y)\,/\,[(y+1)^3\,(y-1)^3] \nonumber \\ 
 & &  -64\,y^3\,H(0,-1,-1;y)\,/\,[(y+1)^3\,(y-1)^3] \nonumber \\ 
 & &  -16\,y^3\,H(0,0,0;y)\,/\,[(y+1)^3\,(y-1)^3] \nonumber \\ 
 & &  +4\,(3\,y^4+2\,y^3+4\,\zeta(2)\,y^2-2\,y-3)\,y\,H(0;y)\, \nonumber \\ 
 & & \phantom{ +4\,(} /\,[(y+1)^3\,(y-1)^3] \nonumber \\ 
 & &  -8\,(3\,y^2+2\,y+3)\,y\,H(-1;y)\,/\,[(y-1)^2\,(y+1)^2] \nonumber \\ 
 & &  -8\,(y^3-5\,y^2-3\,y-1)\,y\,H(0,-1;y)\,/\,[(y-1)^3\,(y+1)^2] \nonumber \\ 
 & &  -8\,(y^2+1)\,y\,H(-1,0;y)\,/\,[(y-1)^2\,(y+1)^2] \nonumber \\ 
 & &  +4\,(y^3-5\,y^2-3\,y-1)\,y\,H(0,0;y)\,/\,[(y-1)^3\,(y+1)^2] \nonumber \\ 
 & &  +16\,(y^2+1)\,y\,H(-1,-1;y)\,/\,[(y-1)^2\,(y+1)^2] \nonumber \\ 
 & &  -4\,(\zeta(2)\,y^4-2\,y^4+4\,y^3+4\,\zeta(2)\,y^3+8\,\zeta(2)\,y^2-8\,\zeta(3)\,y^2\nonumber \\ 
 & & \phantom{ -4\,(}
 -4\,y+4\,\zeta(2)\,y+2-\zeta(2))\,y\,/\,[(y+1)^3\,(y-1)^3],
\eea
and
\bea
\mathrm{Im} \, \mathcal{G}_{2}(s,0,m) & = & 32\,y^3\,H(0,-1;y)\,/\,[(y+1)^3\,(y-1)^3] \nonumber \\ 
 & &  -16\,y^3\,H(0,0;y)\,/\,[(y+1)^3\,(y-1)^3] \nonumber \\ 
 & &  +4\,(y^3-5\,y^2-3\,y-1)\,y\,H(0;y)\,/\,[(y-1)^3\,(y+1)^2] \nonumber \\ 
 & &  -8\,(y^2+1)\,y\,H(-1;y)\,/\,[(y-1)^2\,(y+1)^2] \nonumber \\ 
 & & 
 +4\,(3\,y^4+2\,y^3-4\,\zeta(2)\,y^2-2\,y-3)\,y\,/\,[(y+1)^3\,(y-1)^3].\nonumber \\
\eea
Now to the case of massive internal and massless external quarks. 
An on-shell  two-gluon intermediate state  in 
Fig.~1~(a) could lead for $s>0$ to an imaginary part of 
$G_2$  only (for an off-shell $Z$ boson),
 as a consequence of 
the  Landau-Yang theorem \cite{Landau}. But this form factor is zero
for massless on-shell $q, \bar q$. 
The form factor $G_1$ has an absorptive part only if  $s>4 m^2$.  
In the region  $0 <s < 4m^2$ $G_1$ is real and we get:
\bea
 \mathcal{G}_{1}(s,m,0) & = & 3 \ln(m^2/\mu^2) \, - 9 + \frac{8}{r}
 \zeta(3) +  \frac{24 \phi^2}{r} - \frac{16}{r^2}\zeta(3)
   - \frac{16\phi^2 }{r^2}   + \frac{6 \phi}{w} - 10 \phi^2  \nonumber \\
& &  + \frac{4\phi}{rw} \ln (r) + \frac{16 \phi^2}{r} \ln (r) - \frac{2
  \phi}{w} \ln (r) - \frac{8 \phi}{w}  \ln (4 -r) \nonumber \\
& &            + \frac{4}{rw} Cl_2(2\phi)  + \frac{16 \phi}{r^2}
      Cl_2(2\phi)  + \frac{6}{w} Cl_2(2\phi) + \frac{16\phi}{r}
      Cl_2(4\phi) \nonumber \\ 
& & - \frac{4}{w} Cl_2(4\phi)
          - \frac{16}{r} Cl_3(2\phi)
          + \frac{16}{r^2} Cl_3(2\phi)
          + \frac{8}{r} Cl_3(4\phi)  \, ,
\eea
where $r=s/m^2$, $w = \sqrt{s/(4m^2-s)}$, 
$\phi = \arctan(w)   = \arcsin({\sqrt{r/4}})$, 
and $Cl_2$ ($Cl_3$) denotes the Clausen function of second (third)
order \cite{Lewin}. \\
For $s > 4m^2$ we find:
\bea
 \mathrm{Re} \, \mathcal{G}_{1}(s,m,0) & = & 3\,\ln(s/\mu^2) \, +  
2\,(5\,y^2+8\,y+7)\,y^2\,H(0,0;y)\,/\,(y+1)^4 \nonumber \\ 
 & &  -4\,y^2\,H(0,0,0;y)\,/\,(y+1)^4 \nonumber \\ 
 & &  +8\,y^2\,H(0,-1,0;y)\,/\,(y+1)^4 \nonumber \\ 
 & &  +2\,(3\,y^3+9\,y^2+4\,\zeta(2)\,y^2+10\,\zeta(2)\,y+9\,y+4\,\zeta(2) \nonumber \\ 
 & & \phantom{ +2\,(} +3)\,y\,H(0;y)\,/\,(y+1)^4 \nonumber \\ 
 & &  -16\,y\,H(-1,0,0;y)\,/\,(y+1)^2 \nonumber \\ 
 & &  -16\,y\,H(0,1,0;y)\,/\,(y+1)^2 \nonumber \\ 
 & &  -6\,(y^2-8\,\zeta(2)\,y+2\,y+1)\,H(-1;y)\,/\,(y+1)^2 \nonumber \\ 
 & &  +8\,(y-1)\,H(1,0;y)\,/\,(y+1) \nonumber \\ 
 & &  -2\,(y-1)\,(y^2+1)\,H(-1,0;y)\,/\,(y+1)^3 \nonumber \\ 
 & &  -(9\,y^4+32\,\zeta(2)\,y^4+36\,y^3+56\,\zeta(2)\,y^3-8\,\zeta(3)\,y^3 \nonumber \\ 
 & & \phantom{ -(} +42\,\zeta(2)\,y^2+54\,y^2+36\,y-8\,\zeta(2)\,y-8\,\zeta(3)\,y \nonumber \\ 
 & & \phantom{ -(} +9-2\,\zeta(2))\,/\,(y+1)^4 \, , \label{Rgs4m} 
\eea
and
\bea
\mathrm{Im} \, \mathcal{G}_{1}(s,m,0) & = & 8\,y^2\,H(0,-1;y)\,/\,(y+1)^4 \nonumber \\ 
 & &  -4\,y^2\,H(0,0;y)\,/\,(y+1)^4 \nonumber \\ 
 & &  +2\,(5\,y^2+8\,y+7)\,y^2\,H(0;y)\,/\,(y+1)^4 \nonumber \\ 
 & &  -16\,y\,H(0,1;y)\,/\,(y+1)^2 \nonumber \\ 
 & &  -16\,y\,H(-1,0;y)\,/\,(y+1)^2 \nonumber \\ 
 & &  +8\,(y-1)\,H(1;y)\,/\,(y+1) \nonumber \\ 
 & &  -2\,(y-1)\,(y^2+1)\,H(-1;y)\,/\,(y+1)^3 \nonumber \\ 
 & &  +2\,(3\,y^3+9\,y^2+4\,\zeta(2)\,y^2+9\,y+6\,\zeta(2)\,y+3 \nonumber \\ 
 & & \phantom{ +2\,(} +4\,\zeta(2))\,y\,/\,(y+1)^4 \,  -3 \, .
\eea
For the completely massless case we have for $s>0$:
\be
\mathrm{Re} \,\mathcal{G}_{1}(s,0,0)  =  3\,\ln(s/\mu^2) -9 +
2\,\zeta(2) ,
\label{re_f00_asym}
\ee
and
\be
\mathrm{Im} \, \mathcal{G}_{1}(s,0,0)   =   -3. \label{im_f00_asymptotik}
\ee

\subsection{Check of the Chiral Ward Identities\label{check_WI}}

The anomalous Ward identity (\ref{wardrama}) reads in terms of the
form factors:
\be
 G_{1,R}(s,m,m) +\frac{s}{4m^2}  G_{2,R}(s,m,m)  = 
A_R(s,m,m) + \frac{\alpha_{s}}{4\pi}  T_R \, f_R(s,m) \, ,
\label{warff}
\ee
and Eq. (\ref{wardsama}) yields
\be
 G_{1,R}(s,0,m) +\frac{s}{4m^2}  G_{2,R}(s,0,m) = 
 \frac{\alpha_{s}}{4\pi}  T_R \, f_R(s,m) \, .
\label{warsff}
\ee
The results of Section \ref{subsec_analy} satisfy this equation, both
for spacelike and timelike $s$.
Eqs. (\ref{wardmo}) and  (\ref{qnonsingR})  are trivially satisfied
because $G_{1,R} \bar{u}(p_1)\not \! p \g5 v(p_2) =0$ for massless
external quarks. \\
For the type B
contributions the non-singlet Ward identity (\ref{nonsingR}) has the form:
\be
 \Delta G_{1,R} +\frac{s}{4m^2}  \Delta G_{2,R} =
A_R(s,m,m) \, ,
\label{warffns}
\ee
where  
\be
\Delta G_{j,R} = G_{j,R}(s,m,m) - G_{j,R}(s,0,m) =  G_{j}(s,m,m) -
G_{j}(s,0,m) \, .
\ee
As Eqs. (\ref{warff}) and  (\ref{warsff}) are satisfied by our
 form factors, Eq. (\ref{warffns})  is of course fulfilled, too. 

A specific combination of the above  vertex functions, namely
$G_{1}(s,m,0) - G_{1}(s,0,0)$, which is relevant for
$Z \to$  massless $ b {\bar b}$, was computed in \cite{Kniehl:1989qu}
using Cutkosky rules and dispersion relations. Applying the notation
of that paper, one obtains the relation 
${\cal G}_{1}(s,m,0) - {\cal G}_{1}(s,0,0) = - I_2$ where the function
$I_2$ is given in  \cite{Kniehl:1989qu}
for $s<0$ and $0<s<4 m^2$.  Using in the 
respective kinematical regions
the real and imaginary parts of the form factors
${\cal G}_1$  given above,  an analytical
comparison with Eqs. (2.27), (2.28), (2.30) and (2.31) 
of \cite{Kniehl:1989qu} yields full agreement. 

\section{Threshold and Asymptotic Expansions \label{sec_thr_exp}}

Next  we  expand the  axial vector and pseudoscalar form factors for a massive
external $Q \bar Q$ pair near threshold ($s\sim 4m^2$), in powers of
$\beta = \sqrt{1-\frac{4m^2}{s}}.$
This is useful for applications
to, e.g., $e^+ e^- \to t \bar t$ near threshold. \\
Up to and including terms 
of order $(\beta)^2$ -- respectively of  order $(\beta)^3$ if the
leading term vanishes -- 
the real and imaginary parts are:
\bea
\mathrm{Re}\, \mathcal{G}_{1}(s,m,m) & = &
-\frac{19}{3}\,\zeta(2)-\frac{23}{3}+16\,\zeta(2)\,\ln(2) +
3\,\ln(m^2/\mu^2)\nonumber \\
&& + \beta^2\Bigg[{\frac {11}{15}}\,\zeta(2) -{\frac
  {64}{5}}\,\zeta(2) \ln(2) +\frac{8}{3}\Bigg],\nonumber \\
\mathrm{Im}\, \mathcal{G}_{1}(s,m,m) & = & \beta^3\Bigg[ {\frac {32}{9}}\,\ln(\beta)
+\frac{16}{3}\,\ln(2) - {\frac {44}{9}}\Bigg], \nonumber \\
\mathrm{Re}\, \mathcal{G}_{2}(s,m,m) & = &
-\frac{21}{2}\,\zeta(3)-4\,\zeta(2)\,\ln(2)+\frac{34}{3}\,\zeta(2)+\frac{2}{3}+4\,\ln(2)
 - 12\,\zeta(2)\,\beta\nonumber \\
&& + \beta^2\Bigg[ -6-{\frac {226}{15}}\,\zeta(2)
+\frac{35}{2}\,\zeta(3) +\frac{4}{3} \,\ln(2)
+\frac{44}{5}\,\zeta(2) \ln(2) \Bigg], \nonumber \\
\mathrm{Im}\, \mathcal{G}_{2}(s,m,m) & = & -2+3\,\zeta(2)  + 4\, \beta\, \Bigg[1-\,\ln(2)\Bigg] +
\beta^2\Bigg[-\frac{2}{3}-5\,\zeta(2) \Bigg], \nonumber \\
\mathrm{Re}\, \mathcal{G}_{1}(s,0,m) & = &
\frac{4}{3}\,\zeta(2)-\frac{8}{3}\ln^2(2)+8\,\ln(2)-\frac{23}{3}+3\,\ln(m^2/\mu^2)
\nonumber \\
&& + \beta^2\Bigg[ \frac{8}{15}\,
\ln^2(2)+\frac{38}{15}-\frac{4}{15}\,\zeta(2) \Bigg], \nonumber \\
\mathrm{Im}\, \mathcal{G}_{1}(s,0,m) & = & -4+\frac{8}{3}\,\ln(2) -\frac{8}{15}\,\beta^2\,\ln(2), \nonumber \\
\mathrm{Re}\, \mathcal{G}_{2}(s,0,m) & = &
-\frac{4}{3}\,\zeta(2)+\frac{8}{3}\,\ln^2(2)-4\,\ln(2)+\frac{2}{3} \nonumber\\
&& + \beta^2\Bigg[ -\frac{38}{15}-\frac {16}{5}\,
\ln^2(2)+\frac{16}{3}\,\ln(2) +\frac{8}{5} \,\zeta(2) \Bigg], \nonumber \\
\mathrm{Im}\, \mathcal{G}_{2}(s,0,m) & = & 2-\frac{8}{3}\,\ln(2)\, +
\beta^2\Bigg[ -\frac{8}{3} +\frac{16}{5}\,\ln(2)\Bigg] \, . 
\eea
 The form factor ${\cal G}_{1}(s,m,0)$ is given,
for $r = s/m^2 \to 0$, approximately by  
\be
\mathcal{G}_{1}(s,m,0)  = 3 \ln(m^2/\mu^2) + 
\frac{3}{2}         - \frac{10}{27} r
         - \frac{329}{10800}r^2 \, ,
\label{exg1m0}
\ee
while for $s\sim 4m^2$ its imaginary part has the small $\beta$ expansion
\bea
\mathrm{Im}\, \mathcal{G}_{1}(s,m,0) & = & \Theta (s-4m^2) \,
 \beta^3\Bigg[ {\frac {16}{3}}\,
\ln(\beta)+8\,\ln(2) - {\frac {64}{9}}\Bigg] \, .
\label{exig1m0}
\eea
For brevity we give for the pseudoscalar form factor only the terms of
order $(\beta)^0$:
\bea 
\mathrm{Re} \, A_R(s,m,m) & = &  \left(\frac{\alpha_{s}}{2\pi}\right)^2 \,  C_F\,T_R\;\Big[5\,\zeta(2)
-\frac{21}{2}\,\zeta(3) +12\,\zeta(2) \ln(2) \Big] \, , \nonumber \\
\mathrm{Im} \, A_R(s,m,m) & = &  \left(\frac{\alpha_{s}}{2\pi}\right)^2 \, 3\,\zeta(2)
C_F\,T_R \, .
\eea

Finally we study the behaviour of the form factors involving the
massive quark  for large squared momentum transfer $s\gg m^2$. 
We obtain, up to and including terms of order
$r^{-2}$, with $L=\ln (r)$:
\bea
\mathrm{Re}\, \mathcal{G}_{1}(s,m,m) & = &  3\,\ln(s/\mu^2) -9+2\,\zeta(2)  \nonumber \\
 &&
 +\frac{2}{r}\,(- L^2+(3-4\,\zeta(2))\,L -1+12\,\zeta(2)-8\,\zeta(3)) \nonumber \\
&& +\frac{1}{r^2}\Big[ -2+24\,L+140\,\zeta(2) +32\,L\zeta(3)
-32\,L\zeta(2) -18\,{L}^{2}\nonumber \\
&& \quad \quad +\frac{4\,L^3}{3}-\frac{256}{5}\,\zeta^2(2)+8\,\zeta(2) {L}^{2}+8\, (-2\,L+2)\zeta(2)
-64\,\zeta(3)\Big],\nonumber\\
\mathrm{Im}\, \mathcal{G}_{1}(s,m,m) & = &
-3+\frac{2}{r}\,(2\,L -3+4\,\zeta(2))\nonumber\\
&& +\frac{1}{r^2}\Big[ -24+36\,L-4\,{L}^{2}-32\,\zeta(3)
+32\,\zeta(2) -16\,L\zeta(2)\Big],\nonumber \\
\mathrm{Re}\, \mathcal{G}_{2}(s,m,m) & = &
\frac{2}{r}\,(L^2 -6\,L +4-2\,\zeta(2))\nonumber\\
&& +\frac{1}{r^2}\Big[ -32\,L+4\,{L}^{2}-72\,\zeta(2) -48\,L\zeta(3)
+64\,\zeta(3) \nonumber\\
&& \quad \quad -\frac{116}{5}\, \zeta^2(2)
+\frac{1}{3}\,{L}^{4}-4\,\zeta(2) {L}^{2}+32\,L\zeta(2)\Big],\nonumber \\
\mathrm{Im}\, \mathcal{G}_{2}(s,m,m) & = &
\frac{4}{r}\,(3-L) +\frac{1}{r^2}\Big[32-8\,L+48\,\zeta(3) -8\,L\zeta(2) -\frac{4}{3}\,
{L}^{3}-32\,\zeta(2)\Big],\nonumber \\
\mathrm{Re}\, \mathcal{G}_{1}(s,0,m) & = & 3\,\ln(s/\mu^2)
-9+2\,\zeta(2) +\frac{2}{r}\,(-L^2 +3\,L -1) \nonumber\\
&& +\frac{1}{r^2} \Big[\frac{5}{2}+33\,L-10\,\zeta(2)
+\frac{2}{3}\,{L}^{3}+8\,\zeta(3) -4\,L\zeta(2) -13\,{L}^{2}\Big],\nonumber \\
\mathrm{Im}\, \mathcal{G}_{1}(s,0,m) & = &
-3+\frac{2}{r}\,(2\,L -3)+\frac{1}{r^2}\Big[ -33+26\,L-2\,{L}^{2}-4\,\zeta(2)\Big],\nonumber \\
\mathrm{Re}\, \mathcal{G}_{2}(s,0,m) & = &
\frac{2}{r}\,(L^2 -6\,L +4-2\,\zeta(2)) +\frac{1}{r^2}
(-32\,L+8\,\zeta(2) +12\,{L}^{2}) ,\nonumber \\
\mathrm{Im}\, \mathcal{G}_{2}(s,0,m) & = & \frac{4}{r}\,(3-L) +\frac{1}{r^2}(32-24\,L),\nonumber \\
\mathrm{Re}\, \mathcal{G}_{1}(s,m,0) & = & -9+3\,\ln (s/\mu^2)
+2\,\zeta(2) +\frac{8}{r} \Big[\zeta(3) - \,\zeta(2)\,L\Big]\nonumber\\
&& +\frac{1}{r^2}\Bigg[{\frac {27}{2}}-6\,\zeta(2) +\frac{2}{3}\,{L}^{3}+3\,L-{L}^{2}-
16\,\zeta(3) \nonumber\\
&& \quad \quad +12\,L\zeta(2) +8\, (2-2\,L) \zeta(2)\Bigg],\nonumber \\
\mathrm{Im}\, \mathcal{G}_{1}(s,m,0) & = &
-3+\frac{8\,\zeta(2)}{r}+\frac{1}{r^2}(-3+2\,L-4\,\zeta(2)
-2\,{L}^{2}) \, . \label{asyfofa}
\eea 
The leading asymptotic terms of the pseudoscalar form factor are
\bea 
\mathrm{Re}\, A_{R}(s,m,m) & = &\left(\frac{\alpha_{s}}{2\pi}\right)^2 \, 
 C_F\,T_R\;\frac{1}{r}\big[1/12\,L^4 +4\,\zeta(2)
-{\frac{29}{5}}\,\zeta^2(2) \nonumber \\
& & -2\,L^2 -\zeta(2)
L^2 -12\,L \zeta(3) \big] \,,\nn \\
\mathrm{Im}\, A_{R}(s,m,m) & = &\left(\frac{\alpha_{s}}{2\pi}\right)^2 \,
 C_F\,T_R\;\frac{1}{r}\big[4\,L +12\,\zeta(3) -2\,L \zeta(2)
-1/3\,L^3\big] \, . 
  \label{asyarei}
\eea
The limit $r \to \infty$ corresponds to the massless limit $m \to 0$.
Therefore the chirality-flipping form factors ${\cal G}_2$ and $A_R$
are, as expected, of order $1/r$, and the terms of order $r^0$ in the
above expansion of the chirality-conserving
form factors  ${\cal G}_1$ are  equal to ${\rm Re} \, {\cal G}_1 (s,0,0)$
and ${\rm Im} \, {\cal G}_1 (s,0,0)$, respectively. 

In Figs.~3 -- 6  we have plotted the real and imaginary parts of
the form factors  $C_F T_R {\cal G}_j$ 
as functions of the rescaled center-of-mass energy $\sqrt{s}/(2m)$,
for $\mu = m$ and $N_c=3$.  Also shown are the values of the  asymptotic
expansions of these form factors given in Eq.  (\ref{asyfofa}), and
of the  threshold expansions, Eqs.  (\ref{exg1m0}) and
(\ref{exig1m0}). 
As far as the threshold expansions are
concerned, these Figures show that both  the  next-to-leading terms in
the expansions and restriction to a relatively small region
above threshold  are required  in 
order to get a satisfactory approximation to  the exact result. 
An exception is the expansion of ${\rm Re}\, {\cal G}_1(s,m,0)$ around
$s \sim 0$ where the terms up to second order  approximate this form
factor  very well up to ${\sqrt s} \sim 1.5 m$.
The asymptotic expansions (\ref{asyfofa}), on the other hand, provide
a very good approximation to the form factors over a wide range of 
center-of-mass energies.

\begin{figure}[t!]
\centering
\includegraphics[width=0.47\linewidth]{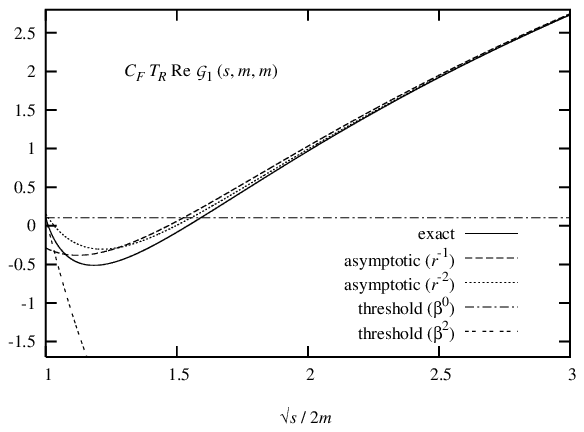}
\includegraphics[width=0.47\linewidth]{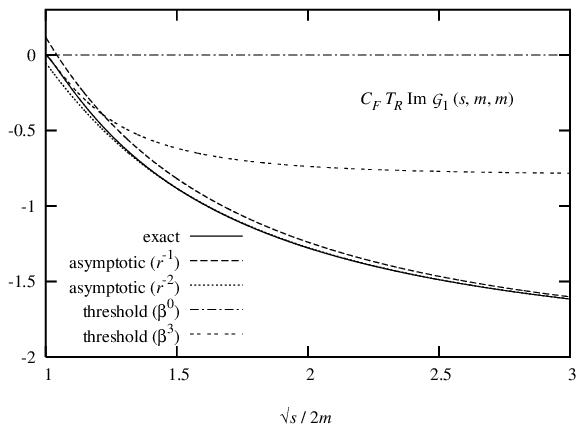}\\
\includegraphics[width=0.47\linewidth]{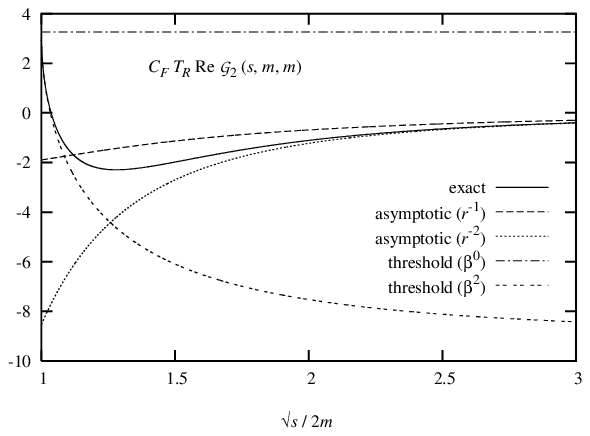}
\includegraphics[width=0.47\linewidth]{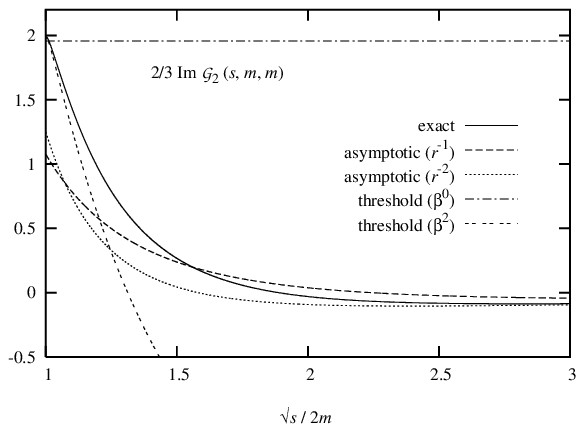}\\[-5pt]
\caption{ The real and imaginary parts of
 $C_F T_R {\cal G}_{1,2}(s,m,m)$ as functions of the rescaled c.m. energy
and their threshold and asymptotic expansions.}
\centering
\includegraphics[width=0.47\linewidth]{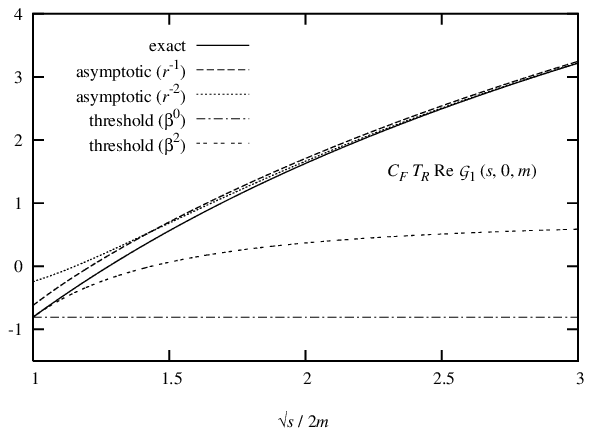}
\includegraphics[width=0.47\linewidth]{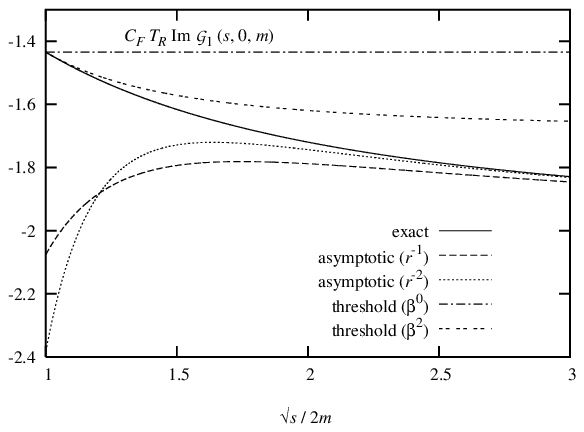}\\
\includegraphics[width=0.47\linewidth]{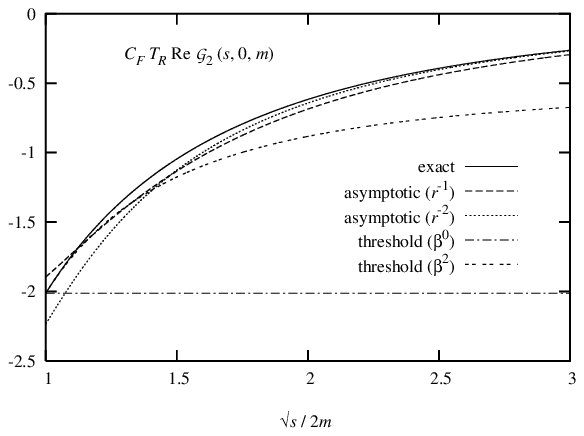}
\includegraphics[width=0.47\linewidth]{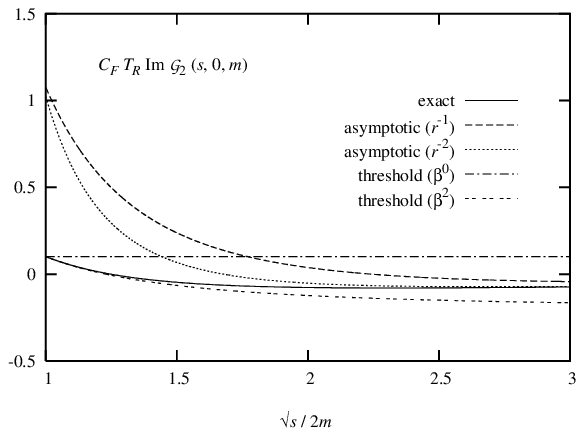}\\[-5pt]
\caption{The real and imaginary parts of $ C_F T_R {\cal G}_{1,2}(s,0,m)$
 as functions of the rescaled c.m. energy
and their threshold and asymptotic expansions.}

\end{figure}
\clearpage

\begin{figure}[th!]
\centering
\includegraphics[width=0.47\linewidth]{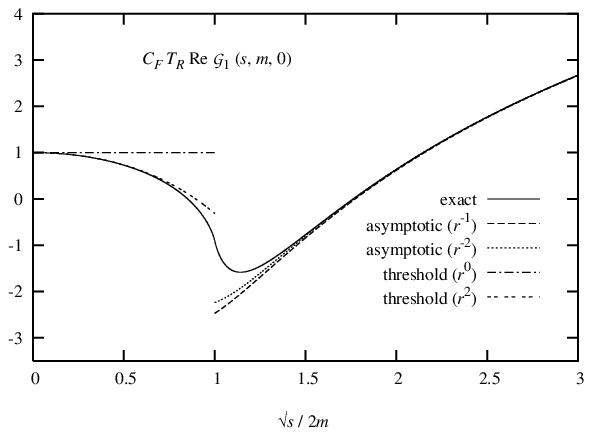}
\includegraphics[width=0.47\linewidth]{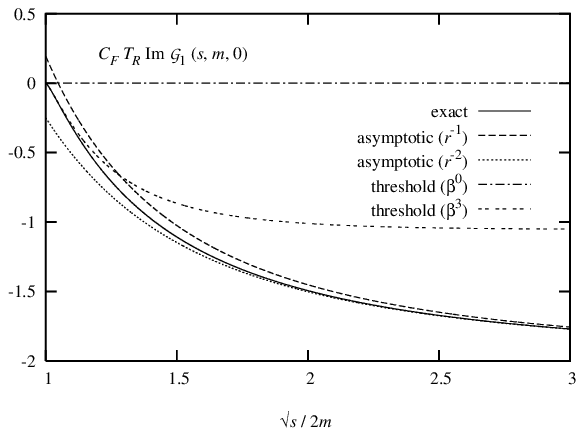}\\[10pt]
\caption{The real and imaginary parts of $C_F T_R {\cal G}_{1}(s,m,0)$
 as functions of the rescaled c.m. energy
and their threshold and asymptotic expansions.}
\centering
\includegraphics[width=0.47\linewidth]{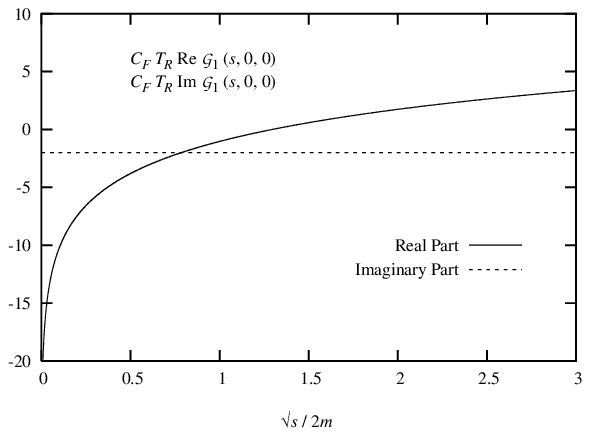}

\caption{The real and imaginary parts of
 $ C_F T_R {\cal G}_{1}(s,0,0)$ as functions of the rescaled c.m. energy.}
\end{figure}

\section{Summary}

In this paper we calculated the triangle diagram contributions 
to the form factors of a heavy-quark axial vector current, of the  flavor-singlet
and non-singlet axial vector currents, and of the corresponding
pseudoscalar currents to second order in the QCD coupling, for four different combinations of
internal and external masses. Moreover, we have given the
threshold and asymptotic expansions of these form factors to a degree
of approximation which should prove useful for applications. 
We compared our exact results with these expansions at threshold
and for large energies. For the threshold expansions we found that
they approximate the exact results only over a very restricted energy range
above threshold, and that at least the second order terms in the expansion
are required for a reliable description. The large energy expansions appear
to converge very well, and do in general yield a good description even
for energies well below the asymptotic regime.

In this  calculation we have used the
implementation of the $\g5$ matrix according to
\cite{Akyeampong:1973xi,Larin:1993} 
in dimensional regularization.
This method is algebraically consistent and convenient for
computations, 
but breaks chiral invariance in $D\neq 4.$  
To restore chiral invariance in this scheme, finite counterterms
are required. With the exception of the counterterm for the pseudoscalar
singlet current, these were determined in \cite{Larin:1993}  
for massless quarks.
In this work, we calculated all finite counterterms required by chiral
invariance in this scheme
for the case for massive quarks, including the counterterm
for the pseudoscalar singlet current (which turns out to be zero). All
counterterms agree with the massless results of \cite{Larin:1993}, 
illustrating their
purely ultraviolet origin. Using these counterterms, we have shown that
the appropriately renormalized vertex functions satisfy the correct
chiral Ward identities.
 This exemplifies for a
non-trivial two-loop case that the counterterms of \cite{Larin:1993}
can be applied, as expected, also to  infrared-finite Green
functions involving massive quarks. However, it is a separate issue  to 
apply  the method  of \cite{Larin:1993} to  Green
functions (with massless and/or massive quarks) that are
infrared-divergent -- using dimensional regularization as well
 -- and ask whether these Green
functions satisfy  the correct chiral Ward identities.
We shall report on this
issue and on applications of our results in this paper and in 
\cite{Bernreuther:2004ih,Bernreuther:2004th} to the  electroproduction
and, in particular, to the forward-backward asymmetry of heavy 
quarks in a future publication.

\section*{Acknowledgement}

This work was supported 
 by Deutsche Forschungsgemeinschaft (DFG), 
SFB/TR9, by DFG-Graduiertenkolleg RWTH Aachen, and by the Swiss 
National Science Foundation (SNF) under contract 200021-101874.


}
\end{fmffile}

\end{document}